\documentclass[twoside]{dis04av}

\usepackage[dvips]{color}
\usepackage{epsfig}
\usepackage{axodraw}

\def\runauthor{A. Vogt, S. Moch and J. Vermaseren}
\def\shorttitle{The three-loop splitting functions in QCD}

\def\z#1{{\zeta_{#1}}}
\def\ca{{C^{}_{\!A}}}
\def\casq{{C^{\, 2}_{\!A}}}
\def\cath{{C^{\, 3}_{\!A}}}

\def\nf{{n^{}_{\! f}}}
\def\n2f{{n^{\,2}_{\! f}}}

\def\S(#1){{{S}_{#1}}}
\def\Ss(#1,#2){{{S}_{#1,#2}}}
\def\Sss(#1,#2,#3){{{S}_{#1,#2,#3}}}
\def\Ssss(#1,#2,#3,#4){{{S}_{#1,#2,#3,#4}}}
\def\Sssss(#1,#2,#3,#4,#5){{{S}_{#1,#2,#3,#4,#5}}}
\def\Npm{{{\bf N_{\pm}}}}
\def\Npmone{{{\bf N_{\pm 1}}}}
\def\Npmi{{{\bf N_{\pm i}}}}
\def\Nminus{{{\bf N_{-}}}}
\def\Nplus{{{\bf N_{+}}}}
\def\Nminustwo{{{\bf N_{-2}}}}
\def\Nplustwo{{{\bf N_{+2}}}}

\def\pqq(#1){p_{\rm{qq}}(#1)}
\def\pqg(#1){p_{\rm{qg}}(#1)}
\def\pgq(#1){p_{\rm{gq}}(#1)}
\def\pgg(#1){p_{\rm{gg}}(#1)}
\def\H(#1){{\rm{H}}_{#1}}
\def\Hh(#1,#2){{\rm{H}}_{#1,#2}}
\def\Hhh(#1,#2,#3){{\rm{H}}_{#1,#2,#3}}
\def\Hhhh(#1,#2,#3,#4){{\rm{H}}_{#1,#2,#3,#4}}

\newcommand{\gsim}{\raisebox{-0.07cm}{$\:\stackrel{>}{{\scriptstyle
 \sim}}\: $} }
\newcommand{\lsim}{\raisebox{-0.07cm}{$\:\stackrel{<}{{\scriptstyle
 \sim}}\: $} }
\newcommand{\beq}{\begin{equation}}
\newcommand{\eeq}{\end{equation}}
\newcommand{\bea}{\begin{eqnarray}}
\newcommand{\eea}{\end{eqnarray}}
\newcommand{\nn}{\nonumber}
\newcommand{\nin}{\noindent}
\newcommand{\MSb}{$\overline{\mbox{MS}}$}
\newcommand{\as}{\alpha_{\rm s}}
\newcommand{\ra}{\rightarrow}
\newcommand{\DD}{{\cal D}}

\begin{document}

\title{The Three-Loop Splitting Functions in QCD}

\author{A. Vogt$^{\,\ast}$, S. Moch$^{\,\dagger}$, 
        J.A.M. Vermaseren$^{\,\ast}$\\[-1mm] \hspace*{1cm} }

\address{$^\ast$NIKHEF,
Kruislaan 409, 1098 SJ Amsterdam, The Netherlands \\
E-mails: avogt@nikhef.nl and t68@nikhef.nl}

\address{$^\dagger$DESY-Zeuthen,
Platanenallee 6, D--15735 Zeuthen, Germany \\
E-mail: moch@ifh.de}

\maketitle

\abstracts{We have computed the next-to-next-to-leading-order (NNLO) 
 contributions to the evolution of unpolarized parton distributions in 
 perturbative QCD [1,2]. In this talk, we briefly recall why this huge 
 computation was necessary and outline how it was performed. We then illustrate 
 the structure of the results and discuss their end-point limits which include 
 the three-loop cusp anomalous dimensions of the Wilson lines. Finally the 
 numerical impact of the new contributions is illustrated.}

\section{Introduction} 

For the next decade, the highest-energy experiments in particle physics will 
be done at the (anti-)proton--proton colliders {\sc Tevatron} and LHC. At such
machines, if we disregard power corrections and observables involving 
final-state fragmentation functions, the cross sections for hard processes $h$ 
can be schematically written as
\beq
\label{eq:pp-fact}
 \sigma_h^{\, pp} \: =\: \sum_{f,f'} f_p \ast f'_p \ast 
 \hat{\sigma}_h^{\,\rm f\:\!f'} \:\: .
\eeq
Here $f_p$ stands for the universal momentum distributions of the partons $f$ 
in the proton, $f = q_i, \bar{q}_i, g$ with $i = 1, \ldots , \nf$, where $\nf$ 
is the number of effectively massless 

\vspace*{-1mm}\noindent
quark flavours. $\hat{\sigma}_h^{\,\rm f\:\!f'}$  
represent the hard (partonic) cross sections for the process under 
consideration. Hence quantitative studies of the standard model, and of  
expected and unexpected new particles, require a precise understanding of the 
partonic luminosities and of the QCD corrections to the corresponding  
cross~sections. 

\vspace*{1mm}
For many important processes, like Higgs-boson production, the 
second-order (NNLO) QCD corrections need to be taken into account, i.e., the 
third term in 
\beq
\label{eq:sig-exp}
 \hat{\sigma}_h^{} \: = \: a_{\rm s}^{n_h}\,
   \left[ \, \hat{\sigma}_h^{(0)} + a^{}_{\rm s}\, \hat{\sigma}_h^{(1)}
   + \, a_{\rm s}^2\, \hat{\sigma}_h^{(2)} + \ldots \,\right] \:\: .
\eeq
The consistent inclusion of $\hat{\sigma}_h^{(2)}$ in Eq.~(\ref{eq:pp-fact})
requires parton distributions evolved with the corresponding 
(process-independent) NNLO splitting functions
\beq
\label{eq:P-exp}
 P_{\:\!\rm f\:\!f'}^{\,\rm NNLO} \: =\:  
     a^{}_{\rm s} P_{\:\!\rm f\:\!f'}^{(0)} 
   + a_{\rm s}^2  P_{\:\!\rm f\:\!f'}^{(1)} 
   + a_{\rm s}^3  P_{\:\!\rm f\:\!f'}^{(2)} \:\: .
\eeq
The one- and two-loop splitting functions have been known for a long time
\cite{Gross:1973rr}--\cite{Hamberg:1992qt}. For the three-loop splitting 
functions $P^{(2)}$, on the other hand, only partial
results had been obtained until recently \cite{Larin:1994vu}--\cite
{Berger:2002sv}. 
However earlier this year we have, finally, computed the complete expressions 
of these functions \cite{Moch:2004pa,Vogt:2004mw}.
\vspace*{1mm}

\section{Outline of the calculation}
\setcounter{equation}{0}

We have derived the NNLO splitting functions by computing the partonic 
structure functions in inclusive deep-inelastic scattering (DIS), 
$\gamma^{\ast}(q)+ f(p) \ra X $ with $ Q^2 \equiv - q^2 > 0$ and $p^2 = 0$, up 
to the third order in the strong coupling $a_{\rm s} = \as /(4\pi)$. This 
computation has been performed for all even or odd values of the Mellin 
variable $N$ via the three-loop forward Compton amplitudes, $\gamma^{\ast}(q) + 
f(p) \ra \gamma^{\ast}(q) + f(p)$. 

\vspace*{1mm}
This approach has two major advantages: Firstly it enables us to obtain, at 
almost the same time, also the three-loop coefficient functions in DIS 
\cite{MVV5}. 
Secondly it allows us to check our programs, at almost any stage, by falling 
back to the {\sc Mincer} program \cite{Gorishnii:1989gt,Larin:1991fz} employed 
in the fixed-$N$ calculations of refs.~\cite{Larin:1994vu}--\cite{Retey:2000nq}. 

\subsection{Mass-factorization in DIS}

Before we address the main computational task, we briefly sketch how the 
splitting functions are extracted from the calculation. We start by writing
the physical structure functions $F_a$ in terms of the (perturbatively 
calculable) bare partonic structure functions $\tilde{F}_{a,k}$, the bare 
coupling $\tilde{a}_{\rm s}$ and the bare parton distributions $\tilde{f}_{k}$, 
\beq
\label{eq:F-bare}
  F_a(Q^2) \: = \: \tilde{F}_{a,k}(\tilde{a}_{\rm s}, Q^2, \varepsilon) 
  \ast \tilde{f}_k \:\: .
\eeq
Summation over the parton species $k$ is understood, and $\ast$ stands for
either the convolution in Bjorken-$x$ space or a simple multiplication of the
Mellin moments.
As indicated in Eq.~(\ref{eq:F-bare}), we use dimensional regularization with
$D = 4 - 2\varepsilon$, thus the singularities of $\tilde{F}_{a,k}$ appear as
poles $\varepsilon^{-l}$. After the ultraviolet divergences have been removed
by coupling-constant renormalization, at the renormalization scale $\mu_r$, 
only initial-state mass singularities remain. They arise when two momenta
become collinear, e.g., $p$ and $k$ in Fig.~1, leading to propagator 
denominators 
$$
  (p-k)^2 \, = \, -2 |\vec{p}\, | |\vec{k}| ( 1- \cos \vartheta) \,
  \stackrel{\vartheta \rightarrow 0}{-\!\!\!-\!\!\!\longrightarrow}
  \, - |\vec{p}\, | |\vec{k}| \vartheta^2 \:\: .
$$
These singularities are removed by mass factorization, at the factorization 
scale~$\mu_f$: $\tilde{F}_{a,k}$ is decomposed into finite pieces, the 
coefficient functions $C_{a,i}$, and the universal transition 
functions $\Gamma_{ik}$ which contain the ($a$-independent) pole parts of 
$\tilde{F}_{a,k}$. The latter are combined with the $\tilde{f}_k$ to form the 
finite renormalized parton densities $f_i$,
\bea
\label{m-fact}
  F_a(Q^2) & \: = \: & \tilde{F}_{a,k} \big(a_{\rm s}(\mu_r^2), 
  \frac{Q^2}{\mu_r^2}, \varepsilon \big) \ast \tilde{f}_k \nn \\[-8mm]
          & \: = \: & { C_{a,i} \big( a_{\rm s}(\mu_r^2),
  \frac{Q^2}{\mu_f^{\,}2},\frac{\mu_f^{\,2}}{\mu_r^2} \big) } \ast 
  \overbrace{ \Gamma_{ik} \big(a_{\rm s}(\mu_r^2), 
  \frac{\mu_f^{\,2}}{\mu_r^2},\varepsilon\big) \ast \tilde{f}_k}^{
  {\textstyle f_i(\mu^2_{\,f_ {\, _{\!  } } }, \mu_r^2)}} \:\: .
\eea
The decomposition (\ref{m-fact}) is not unique. We employ the usual \MSb\ 
scheme where, besides the $1/\varepsilon$ poles, only the artefacts 
$S_{\varepsilon} = \exp( \varepsilon [\,\ln (4\pi) - \gamma_e])$ of dimensional
regularization are removed from the coefficient functions. Differentiation
of Eq.~(\ref{m-fact}) finally leads to the evolution equations for the 
renormalized parton distributions,
\beq
\label{eq:f-evol}
  \frac{\partial}{\partial \ln \mu_f^{\,2}}\, f_i 
  \: = \: \frac{\partial\,\Gamma_{ik}}{\partial \ln\mu_f^{\,2}}\,\ast 
          \tilde{f}_k 
  \: = \: \frac{\partial\,\Gamma_{ik}}{\partial \ln\mu_f^{\,2}}\,\ast
          \Gamma^{-1}_{kj} \ast f_j \:\equiv\: P_{ij} \ast f_j \:\: .
\eeq
The splitting functions (\ref{eq:P-exp}) can thus be obtained 
from the $1/\varepsilon$ poles in Eq.~(\ref{m-fact}).

\begin{figure}[thb]
\label{fig1}
\begin{center}
\begin{picture}(440,100)(-30,0)
\SetWidth{1.0}
   \SetColor{OliveGreen}
   \Gluon(15,12)(38,35){3}{3.5}
   \Line(38,35)(38,65)
   \Line(38,65)(70,65)
   \Photon(38,65)(15,88){-3}{3.5}
   \Line(38,35)(70,35)
   \Gluon(115,12)(138,35){3}{3.5}
   \Line(138,35)(138,65)
   \Line(138,65)(170,65)
   \Photon(138,65)(115,88){-3}{3.5}
   \Line(138,35)(170,35)
   \Gluon(138,50)(170,50){3}{3.5}
   \Gluon(215,12)(238,35){3}{3.5}
   \Line(238,35)(238,65)
   \Line(238,65)(278,65)
   \Photon(238,65)(215,88){-3}{3.5}
   \Line(238,35)(278,35)
   \Gluon(260,35)(260,65){3}{3.5}
   \Text(-2,12)[]{$g(p)$}
   \Text(62,22)[]{$q(k)$}
   \Text(-1,90)[]{$\gamma^\ast (q)$}
\end{picture}
\end{center}
\vspace{-2mm}
\caption{Sample diagrams contributing to the inclusive process 
 $\gamma^{\ast} g \ra X$ up to third order in $a_{\rm s}$.}
\end{figure}
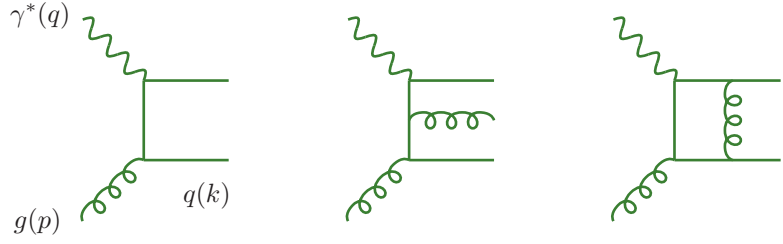

\subsection{The flavour decomposition}

It is convenient to decompose the system (\ref{eq:f-evol}) of $2n_f\!+\!1$
coupled equation as far as possible from charge conjugation and flavour 
symmetry constraints alone. The general structure of the (anti-)quark 
(anti-)quark splitting functions reads
\bea
\label{eq:p-symm}
  P_{{\rm q}_{i}{\rm q}_{k}} \: = \:
  P_{\bar{{\rm q}}_{i}\bar{{\rm q}}_{k}}
  &\! =\! & \delta_{ik} P_{{\rm q}{\rm q}}^{\,\rm v}
        + P_{{\rm q}{\rm q}}^{\,\rm s} \nonumber \\
  P_{{\rm q}_{i}\bar{{\rm q}}_{k}} \: = \:
  P_{\bar{{\rm q}}_{i}{\rm q}_{k}}
  &\! =\! & \delta_{ik} P_{{\rm q}\bar{{\rm q}}}^{\,\rm v}
        + P_{{\rm q}\bar{{\rm q}}}^{\,\rm s}
  \:\: .
\eea
This structure leads to three independently evolving types of flavour 
non-singlet combinations. The flavour asymmetries $q_{\rm ns}^{\,\pm}$ and the 
total valence distribution $q_{\rm ns}^{\rm v}$, 
\beq
\label{eq:q-ns}
  q_{{\rm ns},ik}^{\,\pm} \: = \: q_i^{} \pm
  \bar{q}_i^{} - (q_k^{} \pm \bar{q}_k^{}) \:\: , \quad
  {\textstyle q_{\rm ns}^{\rm v} \: = \: \sum_{r=1}^{\nf} \, (q_r^{}
  - \bar{q}_r^{}) } \:\: ,
\eeq
respectively evolve with
\bea
\label{eq:p-ns}
  P_{\rm ns}^{\,\pm} & \: = \: & P_{{\rm q}{\rm q}}^{\,\rm v}
    \pm P_{{\rm q}\bar{{\rm q}}}^{\,\rm v} \:\: , \nn \\
  P_{\rm ns}^{\,\rm v} & \: = \: & P_{\rm qq}^{\,\rm v}
    - P_{{\rm q}\bar{{\rm q}}}^{\,\rm v} + \nf (P_{\rm qq}^{\,\rm s}
    - P_{{\rm q}\bar{{\rm q}}}^{\,\rm s}) \: \equiv \:
    P_{\rm ns}^{\, -} + P_{\rm ns}^{\,\rm s} \:\: .
\eea
The singlet quark distribution, $q_{\:\!\rm s}^{} \: = \: \sum_{r=1}^{\nf} \, 
(q_r^{} + \bar{q}_r^{})$ is coupled to gluon density $g$,
\beq
\label{eq:ev-sg}
  \frac{d}{d \ln\mu_f^{\,2}}
  \left( \begin{array}{c} \! q_{\:\!\rm s}^{} \! \\ g  \end{array} \right)
  \: = \: \left( \begin{array}{cc} P_{\rm qq} & P_{\rm qg} \\
  P_{\rm gq} & P_{\rm gg} \end{array} \right) \ast
  \left( \begin{array}{c} \!q_{\:\!\rm s}^{}\! \\ g  \end{array} \right)
  \:\: ,
\eeq
where the quark-quark splitting function $P_{\rm qq}$ can be expressed as
\beq
\label{eq:Pqq}
  P_{\rm qq} \: =\: P_{\rm ns}^{\,+} + \nf (P_{\rm qq}^{\:\rm s}
  + P_{\rm {\bar{q}q}}^{\:\rm s})
  \:\equiv\:  P_{\rm ns}^{\,+} + P_{\rm ps}^{} \:\: .
\eeq
The off-diagonal entries in Eq.~(\ref{eq:ev-sg}) are given by
\beq
\label{eq:Poffd}
  P_{\rm qg} \: =\: \nf\, P_{{\rm q}_{i}\rm g} \:\: , \quad
  P_{\rm gq} \: =\: P_{{\rm gq}_{i}}
\eeq
in terms of the flavour-independent splitting functions $P_{{\rm q}_{i} \rm g} 
= P_{\bar{\rm q}_{i}\rm g}$ and $P_{{\rm gq}_{i}}= P_{{\rm g}\bar{\rm q}_{i}}$.

\vspace{1mm}
In the  expansion in powers of $a_{\rm s}$, the flavour-diagonal (`valence') 
quantity $P_{\rm qq}^{\,\rm v}$ in Eq.~(\ref{eq:p-symm}) starts at first order. 
$P_{{\rm q}\bar{{\rm q}}}^{\,\rm v}$ and the flavour-independent (`sea') 
contributions $P_{{\rm qq}}^{\,\rm s}$ and $P_{{\rm q}\bar{{\rm q}}}^{\,\rm s}$
-- and hence the `pure-singlet' term $P_{\rm ps}^{}$ -- are of order 
$\alpha_{\rm s}^2$.
A non-vanishing $P_{\rm ns}^{\,\rm s} = P_{{\rm qq}}^{\,\rm s} - P_{{\rm q}
\bar{{\rm q}}}^{\,\rm s}$ in Eq.~(\ref{eq:p-ns}) occurs for the first time at 
the third order and introduces a new colour structure, $d_{abc}d_{abc}$. 
See Fig.~1b of ref.~\cite{Larin:1991tj} for a typical diagram contributing to
$P_{\rm ns}^{\,\rm s}$.

\subsection{Set-up of the calculation}
Diagrams like those in Fig.~1 have been calculated directly, by working out
the phase-space integrations, for the derivation of the complete second-order 
coefficient functions~\cite{vanNeerven:1991nn}--\cite{Zijlstra:1992qd}. An
extension of this procedure to the third order, however, does not seem
feasible. Instead, we employ the optical theorem
\begin{center}
\SetScale{1.0}
\SetWidth{1.0}
\begin{picture}(420,100)(10,-15)
\SetColor{OliveGreen}
     \Line(100,49)(130,58)
     \Line(100,43)(132,46)
     \Line(100,37)(132,34)
     \Line(100,31)(130,22)
     \Photon(60,70)(95,50){4}{5}
     \Line(60,10)(95,30)
     \COval(100,40)(20,10)(0){Black}{Yellow}
     \Text(43,10)[]{$f(p)$}
     \Text(43,72)[]{$\gamma^{\ast}(q)$}
 
     \Line(260,49)(290,49)
     \Line(260,43)(290,43)
     \Line(260,37)(290,37)
     \Line(260,31)(290,31)
     \Photon(220,70)(255,50){4}{5}
     \Line(220,10)(255,30)
     \COval(260,40)(20,10)(0){Black}{Yellow}
     \Text(211,10)[]{$f$}
     \Text(211,72)[]{$\gamma^{\ast}$}
     \Photon(340,70)(305,50){-4}{5}
     \Line(340,10)(305,30)
     \COval(300,40)(20,10)(0){Black}{Yellow}
     \Text(350,10)[]{$f$}
     \Text(353,72)[]{$\gamma^{\ast}$}
\SetColor{Black}
     \Line(26,0)(26,80)
     \Line(140,0)(140,80)
     \Text(152,79)[]{2} 
     \Text(180,40)[]{$\longleftrightarrow$}
\end{picture}
\end{center}
to transform the problem into forward Compton amplitudes.
We then make use of a theorem~\cite{Ybook} that the coefficient of 
$(2p\cdot q)^N$ provides the $N$-th Mellin moment, 
$$
  \tilde{F}(N) \: = \: \int_0^1 \! dx \: x^{N-1} \tilde{F}(x) \:\: ,
$$
of the partonic structure functions (\ref{eq:F-bare}) which we need to 
calculate.

\vspace{1mm}
In order to obtain the complete set of the third-order contributions to the
splitting functions (\ref{eq:p-ns}) and (\ref{eq:ev-sg}) we have to include,
besides the photon shown above, also the $W$-boson \cite{Larin:1991tj} -- for 
accessing $P_{\rm ns}^{\, -}$ and $P_{\rm ns}^{\,\rm s}$ -- and a fictitious 
classical scalar $\phi $ coupling directly only to the gluon field via 
$\phi\, G_{\mu\nu}^{\,a}G_a^{\,\mu\nu}$ \cite{ZG2a,ZG2b} -- for accessing 
$P_{\rm gq\,}$ and $P_{\rm gg}$. Especially the latter leads to a substantial 
increase of the number of diagrams (generated with {\sc Qgraf}
\cite{Nogueira:1991ex}) as shown in Table 1.
Among the partons $f$ we also include the standard ghost~$h$. This allows us 
to take the sum over external gluon spins by contracting with $-g_{\mu\nu}$ 
instead of the full physical expression. 

\begin{table}[bht]
\vspace*{-4mm}
\[ \begin{tabular}{l c c c c}
\hline\\ & & & & \\[-7mm] 
{} &{tree} &1-loop &2-loop &3-loop \\[1mm]
\hline\\ & & & & \\[-7mm]
q$\gamma$ & 1  & 3  & \phantom025         & \phantom0359 \\
g$\gamma$ & {} & 2  & \phantom017         & \phantom0345 \\
h$\gamma$ & {} & {} & \phantom0\phantom02 & \phantom0\phantom056 \\
q$W$      & 1  & 3  & \phantom032         & \phantom0589 \\
q$\phi$   & {} & 1  & \phantom023         & \phantom0696 \\
g$\phi$   & 1  & 8  & 218                 & 6378 \\
h$\phi$   & {} & 1  & \phantom033         & 1184 \\[1mm]
\hline\\ & & & & \\[-6.5mm]
sum       & 3  & 18 & 350                 & 9607 \\[1mm] 
\hline\\
\end{tabular} \]
\vspace*{-7mm}
\caption{The number of diagrams employed in our calculation of the 
 three-loop splitting functions.}
\vspace*{-4mm}
\end{table}

Obviously a highly efficient symbolic treatment is needed to cope with the
task at hand. Unsurprisingly, we use FORM for all manipulations. Note that the 
capabilities of this program had to be extended substantially for 
this computation~\cite{Vermaseren:2000nd,Vermaseren:2002rp}.

\subsection{Treatment of the integrals}

Finally we illustrate the computation of the integrals required to evaluate the 
forward Compton amplitudes. One of the 9607 three-loop diagrams in Table 1 is 
shown here together with a useful pictorial representation of its 
momentum~flow:

\begin{center}
\SetScale{0.8}
\SetWidth{1.5}
\begin{picture}(300,90)(-30,-25)

\SetColor{Red}  

\Line(35,0)(0,0)
\Line(70,0)(35,0)
\Line(105,0)(70,0)
\Gluon(-15,-15)(0,0){3}{2}
\Gluon(105,0)(120,-15){3}{2}

\SetColor{OliveGreen} 
\Gluon(70,0)(70,60){4}{5} 
\Gluon(35,0)(35,60){4}{5} 

\Photon(105,60)(120,70){2}{2}
\Photon(0,60)(-15,70){2}{2}

\Vertex(0,0){2}      
\Vertex(0,60){2}      
\Vertex(105,0){2} 
\Vertex(35,0){2}
\Vertex(70,0){2}
\Vertex(32,60){2}  
\Vertex(70,60){2}  
\Vertex(105,60){2}


\Line(0,0)(0,60)
\Line(0,60)(35,60)
\Line(35,60)(70,60)
\Line(70,60)(105,60)
\Line(105,60)(105,0)

       \Line(250,55)(300,55) \Line(250,5)(300,5)
       \Line(250,55)(250,5) \Line(300,55)(300,5)
       \CArc(250,30)(25,90,270) \CArc(300,30)(25,270,90)
       \Line(225,30)(215,30) \Line(325,30)(335,30)
{\SetColor{Red} \SetWidth{4} \Line(300,55)(250,55) 
 \CArc(250,30)(25,90,135)\CArc(300,30)(25,45,90) 
 }

\SetColor{Black}
\SetScale{1.5}
\SetWidth{0.8}
\LongArrow(77,16)(97,16)

\end{picture}
\end{center}
\vspace{-2mm}

\nin
For the latter we temporarily disregard the external parton lines and draw the 
remaining self-energy type diagram, the topology of which is denoted following 
the notation of ref.~\cite{Larin:1991fz}. Our example is a ladder (LA) diagram.
The (partly additional) denominators carrying the parton 
momentum $p$ are then indicated by the fat (in the coloured version: red) 
lines. Here $p$ runs, after turning the diagram upside-down, through the lines 
1, 2 and 3, thus the example is assigned the subtopology LA$_{13}$. 

\vspace{1mm}
According to our discussion in the previous subsection, we need analytic 
expressions for the (dimensionless) coefficients $I(N)$ of $(2p\cdot q)^N /
Q^{2\alpha}$. One might try to obtain $I(N)$ by brute force, Taylor-expanding 
the denominators with $p$ and working out the sums. It turns out that such a 
strategy, in general, does not work. Instead, we employ identities based on 
integration by parts, scaling arguments and form-factor decompositions (see 
Sect.~2 of ref.~\cite{Moch:2004pa}) to successively simplify the integrals.

\vspace{1mm}
The LA$_{13}$ integrals, e.g., can be simplified by applying $p^{\mu}\partial / 
\partial q^{\mu}$ both inside and outside the integral. For the scalar integral 
with unit denominators this yields
 
\vspace*{1.1cm}
\raisebox{-34pt}{  
\SetScale{1.0} 
\begin{picture}(50,20)(-10,-5)
\SetColor{OliveGreen} 
\SetScale{0.75}
\SetWidth{1.6}
       \Line(35,70)(85,70) \Line(35,20)(85,20) 
       \Line(35,70)(35,20) \Line(85,70)(85,20) 
        \CArc(35,45)(25,90,270) \CArc(85,45)(25,270,90)
        \Line(10,45)(0,45) \Line(110,45)(120,45)
        {\SetColor{Red} \SetWidth{4} 
        \CArc(35,45)(25,90,135) \CArc(85,45)(25,45,90)
        \Line(35,70)(85,70)} 
\SetColor{Black}
        \PText(60,81)(0)[l]{1} 
        \PText(60,13)(0)[l]{1} 
        \PText(40,45)(0)[l]{1} \PText(80,45)(0)[r]{1}
        \PText(5,60)(0)[l]{1}\PText(20,72)(0)[lb]{1}
        \PText(115,60)(0)[r]{1}\PText(100,72)(0)[rb]{1}
        \PText(15,20)(0)[l]{1} \PText(105,20)(0)[r]{1}
\Text(120,33)[]{=} 
\Text(316,33)[]{(2.10)} 
\end{picture} 
}
\addtocounter{equation}{1}

\vspace*{-2mm}
$$ \hspace*{-8mm}
- \frac{N\!+\!3\!+\!3\epsilon}{N\!+\!2}\, \frac{2p\!\cdot\! q}{q^2}
\hspace*{-2mm}
\raisebox{-34pt}{  
\SetScale{1.0}
\SetColor{OliveGreen} 
\begin{picture}(50,20)(0,-3)
\SetScale{0.75}
\SetWidth{2}
       \Line(35,70)(85,70) \Line(35,20)(85,20) 
       \Line(35,70)(35,20) \Line(85,70)(85,20) 
        \CArc(35,45)(25,90,270) \CArc(85,45)(25,270,90)
        \Line(10,45)(0,45) \Line(110,45)(120,45)
        {\SetColor{Red} \SetWidth{4} 
        \CArc(35,45)(25,90,135) \CArc(85,45)(25,45,90)
          \Line(35,70)(85,70)} 
\SetColor{Black}
        \PText(60,81)(0)[l]{1} 
        \PText(60,13)(0)[l]{1} 
        \PText(40,45)(0)[l]{1} \PText(80,45)(0)[r]{1}
        \PText(5,60)(0)[l]{1}\PText(20,72)(0)[lb]{1}
        \PText(115,60)(0)[r]{1}\PText(100,72)(0)[rb]{1}
        \PText(15,20)(0)[l]{1} \PText(105,20)(0)[r]{1}
\end{picture} 
}
\hspace*{1.8cm}\!\!
+ \frac{2}{N\!+\!2}
\raisebox{-34pt}{  
\SetScale{1.0} 
\begin{picture}(50,20)(0,-3)
\SetScale{0.75}
\SetWidth{2}
\SetColor{OliveGreen} 
       \Line(35,70)(85,70) \Line(35,20)(85,20) 
       \Line(35,70)(35,20) \Line(85,70)(85,20) 
        \CArc(35,45)(25,90,270) \CArc(85,45)(25,270,90)
        \Line(10,45)(0,45) \Line(110,45)(120,45)
        {\SetColor{Red} \SetWidth{4} 
        \CArc(35,45)(25,90,135) 
          \Line(35,70)(85,70)} 
\SetColor{Black}
        \PText(60,81)(0)[l]{1} 
        \PText(60,13)(0)[l]{1} 
        \PText(40,45)(0)[l]{1} \PText(80,45)(0)[r]{1}
        \PText(5,60)(0)[l]{1}\PText(20,72)(0)[lb]{1}
        \PText(105,70)(0)[rb]{2}
        \PText(15,20)(0)[l]{1} \PText(105,20)(0)[r]{1}
        \end{picture} 
}
$$
Here the LA$_{13}$ integral occurs twice, once with a prefactor $2p\cdot q$.
Hence Eq.~(2.10) represents a difference equation (here of order $n=1$) which 
expresses its coefficient $I(N)$ in terms of that of a LA$_{12}$ integral with 
an enhanced denominator in the~3-line,
\beq
\label{eq:diff}
  a_0(N) I(N) - \dots  - a_n(N) I(N\! -\! n)  - G(N) = 0 \:\: .
\eeq
First-order recursion relations like Eq.~(2.10) can be reduced to a sum.
Higher-order recursions (we have used equations up to $n=4$) can be solved by
inserting a suitable ansatz into Eq.~(\ref{eq:diff}). Both procedures make use
of the fact that all integrals required for the computation of the splitting
functions can be expressed in terms of harmonic sums~\cite{Vermaseren:1998uu}.
Recall that these sums are recursively defined by
\beq
  S_{\pm m}(M) \: = \:\sum_{i=1}^{M}\: \frac{(\pm 1)^i}{i^{\, m}} 
  \:\: , \quad
  S_{\pm m_1,m_2,\ldots,m_k}(M) \: = \:  \sum_{i=1}^{M}\:
  \frac{(\pm 1)^{i}}{i^{\, m_1}}\: S_{m_2,\ldots,m_k}(i) \:\: .
\eeq
To the accuracy in the dimensional offset $\varepsilon$ required for the 
calculation of the splitting functions, our example integral reads, using the 
{\sc Form} notations {\tt den(i+N)} for $1/(N\! +\! i)$ and 
{\tt S(R(m1,...,mk),i+N}) for $S_{m_1,\ldots,m_k}(N\! +\! i)$,  

\SetScale{0.8}
\begin{picture}(120,70)(10,0)
\SetColor{OliveGreen}
\SetWidth{2}
       \Line(35,70)(85,70) \Line(35,20)(85,20)
       \Line(35,70)(35,20) \Line(85,70)(85,20)
        \CArc(35,45)(25,90,270) \CArc(85,45)(25,270,90)
        \Line(10,45)(0,45) \Line(110,45)(120,45)
        {\SetColor{Red} \SetWidth{4}
        \CArc(35,45)(25,90,135) \CArc(85,45)(25,45,90)
        \Line(35,70)(85,70)}
\SetColor{Black}
        \PText(60,80)(0)[l]{1}
        \PText(60,14)(0)[l]{1}
        \PText(40,45)(0)[l]{1} \PText(80,45)(0)[r]{1}
        \PText(5,60)(0)[l]{1}\PText(20,72)(0)[lb]{1}
        \PText(115,60)(0)[r]{1}\PText(100,72)(0)[rb]{1}
        \PText(15,19)(0)[l]{1} \PText(105,19)(0)[r]{1}
\Text(125,36)[]{=}
\Text(342,36)[]{(2.13)}
\end{picture}

\vspace{-1mm}
\small
\begin{verbatim}
 +theta(N)*sign(N)*ep^-2*(-8/3*den(1+N)^2-4*den(1+N)^3+8/3*den(1+N)^2*S(R(
    1),1+N)+4/3*den(1+N)*S(R(1),1+N)+2/3*den(1+N)*S(R(2),1+N)-4/3*den(2+N)
    ^2-2*den(2+N)^3+4/3*den(2+N)^2*S(R(1),2+N)+4/3*den(2+N)*S(R(1),2+N)+2/
    3*den(2+N)*S(R(2),2+N)+4/3*S(R(1),N)+2/3*S(R(1,2),N)-2*S(R(2),N)-4/3*
    S(R(2),N)*N+4*S(R(2,1),N)+4/3*S(R(2,1),N)*N-6*S(R(3),N)-2*S(R(3),N)*N)

 +theta(N)*sign(N)*ep^-1*(32*den(1+N)^2+164/3*den(1+N)^3+24*den(1+N)^4-20/
    3*den(1+N)^3*S(R(1),1+N)-88/3*den(1+N)^2*S(R(1),1+N)+8/3*den(1+N)^2*S(
    R(1,1),1+N)-40/3*den(1+N)^2*S(R(2),1+N)-16*den(1+N)*S(R(1),1+N)+8/3*
    den(1+N)*S(R(1,1),1+N)+10/3*den(1+N)*S(R(1,2),1+N)-58/3*den(1+N)*S(R(2
    ),1+N)+10*den(1+N)*S(R(2,1),1+N)-18*den(1+N)*S(R(3),1+N)+16*den(2+N)^2
    +82/3*den(2+N)^3+12*den(2+N)^4-10/3*den(2+N)^3*S(R(1),2+N)-44/3*den(2+
    N)^2*S(R(1),2+N)-6*den(2+N)^2*S(R(2),2+N)-16*den(2+N)*S(R(1),2+N)+8/3*
    den(2+N)*S(R(1,1),2+N)+10/3*den(2+N)*S(R(1,2),2+N)-46/3*den(2+N)*S(R(2
    ),2+N)+6*den(2+N)*S(R(2,1),2+N)-12*den(2+N)*S(R(3),2+N)-20*S(R(1),N)+8/
    3*S(R(1,1),N)+10/3*S(R(1,1,2),N)-16*S(R(1,2),N)-4*S(R(1,2),N)*N+14*S(
    R(1,2,1),N)+4*S(R(1,2,1),N)*N-24*S(R(1,3),N)-6*S(R(1,3),N)*N+56/3*S(R(
    2),N)+20*S(R(2),N)*N-134/3*S(R(2,1),N)-56/3*S(R(2,1),N)*N+16/3*S(R(2,1
    ,1),N)+8/3*S(R(2,1,1),N)*N-62/3*S(R(2,2),N)-22/3*S(R(2,2),N)*N+76*S(R(
    3),N)+100/3*S(R(3),N)*N-10*S(R(3,1),N)-10/3*S(R(3,1),N)*N+36*S(R(4),N)
    +12*S(R(4),N)*N)  .
\end{verbatim}
\normalsize

\vspace{3mm}
Despite being uncharacteristically simple in both derivation and size, 
Eq.~(2.10) illustrates the strict hierarchy of subtopologies in our procedure. 
Our LA$_{13}$ example can only be evaluated once the LA$_{12}$ integral in 
Eq.~(2.10) is known. This integral, in turn, requires the so-called basic 
building blocks (with only one $p$-dependent denominator) LA$_{11}$ and 
LA$_{22}$ together with other integrals of simpler topologies where one of 
the non-$p$ denominators has been removed. Also those integrals need to be 
evaluated in terms of yet simpler cases, and so on.
 
\vspace{1mm}
Constructing the reduction chains for all subtopologies, and computing all
integrals required for evaluating either diagrams or other, higher-level 
integrals took literally years of both human and computing resources. It would 
not have been possible to get through without extensive tabulation of 
intermediate results for which new features were added to {\sc Form}
\cite{Vermaseren:2002rp}. At the end, a database had been accumulated of more 
than 100$\,$000 integrals requiring about 3.5 GBytes of disk~space.

\section{Sample results in {\boldmath{$N$}}-space and {\boldmath{$x$}}-space}
\setcounter{equation}{0}

We illustrate our final results by writing down the even-$N$ anomalous 
dimensions and the corresponding $x$-space splitting functions in 
Quantum-Gluodynamics, i.e., for QCD with $\nf = 0$ quark flavours. The complete 
QCD results in refs.~\cite{Moch:2004pa,Vogt:2004mw} are considerably longer, by 
a factor of about 15, but not structurally more complicated.

\subsection{Expressions in Mellin-$N$ space}

We start in $N$-space where, as discussed above, the actual calculations have 
been performed. Recall that we expand in terms of $a_{\rm s} = \as /(4\pi)$, 
and that $\gamma^{\,(n)}(N) \! = ! - P^{\,(n)}(N)$ is the coefficient of 
$a_{\rm s}^{\, n+1}$. Here we hide $N$-dependent denominators by using 
differences of harmonic sums at suitably shifted arguments for which we employ 
the abbreviations
$$
 \Npmi\, S_{\vec{m}} \equiv S_{\vec{m}}(N \pm i) \:\: , \quad 
 \Npm \equiv \Npmone \:\: .
$$
In this notation the well-known one- and two-loop results~\cite
{Gross:1973rr,Georgi:1974sr,Floratos:1979ny,Floratos:1981hs,Hamberg:1992qt} 
are given by
\small
\small
\bea
  &&\gamma^{\,(0)}_{\,\rm gg}(N) \: = \:
          \ca  \, \*  \Big(
            4 \, \* ( \Nminustwo
          - 2 \* \Nminus
          - 2 \* \Nplus
          + \Nplustwo
          + 3 ) \, \* \S(1)
          - {11 \over 3}
          \Big)
  \label{eq:ggg0} \:\: , \\ &&\gamma^{\,(1)}_{\,\rm gg}(N) \: = \:
       4\,  \*  \casq  \*  \Big(
          - 4\, \* (\Nminustwo-2\*\Nminus-2\*\Nplus+\Nplustwo+3) \* \Big[
            \Ss(1,-2)
          + \Ss(1,2)
          + \Ss(2,1)
       \Big]
  \nonumber\\&& \mbox{}
          + {8 \over 3} \* (\Nplus-\Nplustwo) \* \S(2)
          - 4 \* (\Nminus-3\*\Nplus+\Nplustwo+1) \* \Big[
            3 \* \S(2)
          - \S(3)
       \Big]
          + {109 \over 18} \* (\Nminus+\Nplus) \* \S(1)
  \nonumber\\&& \mbox{}
          + {61 \over 3} \* (\Nminus -\Nplus) \* \S(2)
          - {8 \over 3}
          + 2 \* \S(-3)
          - {14 \over 3} \* \S(1)
          + 2 \* \S(3)
          \Big)
\:\: . \label{eq:ggg1}
\eea
\normalsize
The three-loop gluon-gluon anomalous dimension \cite{Vogt:2004mw} reads, for 
$n_f=0$, 
\small
\bea
  &&\gamma^{\,(2)}_{\,\rm gg}(N) \: = \:
       16\,  \*   \cath  \*  \Big(
	  (\Nminustwo-2\*\Nminus-2\*\Nplus+\Nplustwo+3) \* \Big[
            {73091 \over 648} \* \S(1)
          - 16 \* \Ss(1,-4)
          + {88 \over 3} \* \Ss(1,-3)
  \nonumber\\&& \mbox{}
          + 16 \* \Sss(1,-3,1)
          + {85 \over 6} \* \Ss(1,-2)
          + 4 \* \Sss(1,-2,-2)
          - 11 \* \Sss(1,-2,1)
          + 4 \* \Sss(1,-2,2)
          - {413 \over 108} \* \Ss(1,1)
          + 24 \* \Sss(1,1,-3)
  \nonumber\\&& \mbox{}
          + 11 \* \Sss(1,1,-2)
          - 16 \* \Ssss(1,1,-2,1)
          + 8 \* \Sss(1,1,3)
          - {67 \over 9} \* \Ss(1,2)
          + 8 \* \Sss(1,2,-2)
          + 8 \* \Sss(1,2,2)
          + {55 \over 3} \* \Ss(1,3)
          + 8 \* \Sss(1,3,1)
  \nonumber\\&& \mbox{}
          - 8 \* \Ss(1,4)
          - {395 \over 27} \* \S(2)
          - 14 \* \Ss(2,-3)
          - {11 \over 3} \* \Ss(2,-2)
          + 8 \* \Sss(2,-2,1)
          - {67 \over 9} \* \Ss(2,1)
          + 4 \* \Sss(2,1,-2)
          + 8 \* \Sss(2,1,2)
  \nonumber\\&& \mbox{}
          + {22 \over 3} \* \Ss(2,2)
          + 8 \* \Sss(2,2,1)
          - 10 \* \Ss(2,3)
          + 8 \* \Sss(3,1,1)
          - 8 \* \Ss(3,2)
  	  \Big]
	+ (2-\Nminus-\Nplus) \* \Big[
            {713 \over 324} \* \S(1)
          + {26 \over 3} \* \Ss(1,-3)
  \nonumber\\&& \mbox{}
          - 14 \* \Sss(1,-2,1)
          + {61 \over 9} \* \Ss(1,-2)
          + {80 \over 27} \* \Ss(1,1)
          - 14 \* \Sss(1,1,-2)
          + {109 \over 18} \* \Ss(1,2)
          - 4 \* \Ss(1,3)
  	  \Big]
	+ (\Nminus-\Nplus) \* \Big[
            {473 \over 216} \* \S(2)
  \nonumber\\&& \mbox{}
          - 12 \* \Ss(2,-3)
          + 5 \* \Ss(2,-2)
          - 2 \* \Ss(2,1)
          - 8 \* \Sss(2,1,-2)
          + {23 \over 3} \* \Ss(2,2)
          - 10 \* \Ss(2,3)
          + {665 \over 36} \* \S(3)
          - 20 \* \Ss(3,-2)
          + {34 \over 3} \* \Ss(3,1)
  \nonumber\\&& \mbox{}
          - 16 \* \Ss(3,2)
          - 21 \* \S(4)
          - 26 \* \Ss(4,1)
	\Big]
	+ (\Nminus-\Nplustwo) \* \Big[
          - {9533 \over 108} \* \S(2)
          + 8 \* \Ss(2,-3)
          - {77 \over 3} \* \Ss(2,-2)
          - 8 \* \Sss(2,-2,1)
  \nonumber\\&& \mbox{}
          - 8 \* \Sss(2,1,-2)
          - {44 \over 3} \* \Ss(2,2)
          - {1517 \over 18} \* \S(3)
          + 8 \* \Ss(3,-2)
          - {121 \over 3} \* \Ss(3,1)
          + 4 \* \Ss(3,2)
          + 44 \* \S(4)
          + 16 \* \Ss(4,1)
          - 8 \* \S(5)
	\Big]
  \nonumber\\&& \mbox{}
	+ (1-\Nplus) \* \Big[
            {8533 \over 108} \* \S(2)
          - 8 \* \Ss(2,-3)
          + {103 \over 3} \* \Ss(2,-2)
          + 8 \* \Sss(2,-2,1)
          + {109 \over 9} \* \Ss(2,1)
          + 8 \* \Sss(2,1,-2)
          + {28 \over 3} \* \Ss(2,2)
  \nonumber\\&& \mbox{}
          + {1579 \over 18} \* \S(3)
          + 8 \* \Ss(3,-2)
          + {71 \over 3} \* \Ss(3,1)
          - 4 \* \Ss(3,2)
          - {98 \over 3} \* \S(4)
          - 16 \* \Ss(4,1)
          + 36 \* \S(5)
	\Big]
          - {79 \over 32}
          + 4 \* \S(-5)
          - 8 \* \Ss(-4,1)
  \nonumber\\&& \mbox{}
          + {67 \over 9} \* \S(-3)
          - 4 \* \Ss(-3,-2)
          - 2 \* \Ss(-3,2)
          - 4 \* \Ss(-2,-3)
          - {11 \over 3} \* \Ss(-2,-2)
          + 4 \* \Sss(-2,-2,1)
          + 4 \* \Sss(-2,1,-2)
  \nonumber\\&& \mbox{}
          - {16619 \over 162} \* \S(1)
          - {88 \over 3} \* \Ss(1,-3)
          - {523 \over 18} \* \Ss(1,-2)
          + 11 \* \Sss(1,-2,1)
          + {413 \over 108} \* \Ss(1,1)
          - 11 \* \Sss(1,1,-2)
          - {67 \over 9} \* \Ss(1,2)
  \nonumber\\&& \mbox{}
          - {33 \over 2} \* \Ss(1,3)
          + {781 \over 54} \* \S(2)
          - 4 \* \Ss(2,-3)
          + {11 \over 3} \* \Ss(2,-2)
          + 4 \* \Sss(2,-2,1)
          - {67 \over 9} \* \Ss(2,1)
          + 4 \* \Sss(2,1,-2)
          - {22 \over 3} \* \Ss(2,2)
  \nonumber\\&& \mbox{}
          + {67 \over 9} \* \S(3)
          - 4 \* \Ss(3,-2)
          + {11 \over 6} \* \Ss(3,1)
          - 2 \* \Ss(3,2)
          - 8 \* \Ss(4,1)
          + 4 \* \S(5)
          \Big)
 \:\: . \label{eq:ggg2}
\eea
\normalsize
Note that harmonic sums up to weight $2n+1$ occur at order $\as^{\,n+1}$
(N$^n$LO).

\vspace{1mm}
We stress that Eqs.~(\ref{eq:ggg1}) and (\ref{eq:ggg2}) are directly applicable 
only for even positive values of $N$, while the lowest-order expression 
(\ref{eq:ggg0}) holds for any positive integer. For general (non-integer) $N$, 
$\gamma^{\,(n)}(N)$ can be obtained by numerically evaluating
\beq
\label{eq:Pdef}
  \gamma^{\,(n)}(N) \: = \:
  - \int_0^1 \!dx\:\, x^{\,N-1}\, P^{\,(n)}(x)
\eeq
using the $x$-space results to which we turn now.

\subsection{Expressions in Bjorken-$x$ space}

There is a theorem \cite{Ybook} ensuring that the splitting functions 
$P^{(n)}(x)$ can be uniquely reconstructed from their even-$N$ (or odd-$N$) 
moments obtained in our calculations. In fact, the close relation between the 
harmonic sums and the harmonic polylogarithms facilitates an algebraic 
procedure \cite{Remiddi:1999ew,Moch:1999eb} for the inverse Mellin transform.  

\vspace{1mm}
For a compact representation of the gluon-gluon splitting functions we use
$$
  p_{\rm{gg}}(x) \: \equiv \: (1-x)^{-1} + x^{\,-1} - 2 + x - x^{\,2} 
$$
and an abbreviation for the harmonic polylogarithms~\cite{Remiddi:1999ew},
$$
  H_{\pm (m+1),\,\pm (n+1),\, \ldots} \: \equiv \:
  H_{\footnotesize{\underbrace{0,\ldots ,0}_{\scriptstyle m} },\,
  \pm 1,\, {\footnotesize\underbrace{0,\ldots ,0}_{\scriptstyle n} },
  \, \pm 1,\, \ldots}(x) \:\: .
$$
The one- and two-loop results \cite{Altarelli:1977zs,Furmanski:1980cm} for 
$\nf = 0$ can then be written as
\small
\bea
  &&P^{\,(0)}_{\rm gg}(x) \: = \:
          \ca  \*  \Big(
            4 \* \pgg(x)
          + {11 \over 3} \* \delta(1 - x)
          \Big)
\label{eq:Pgg0}
  \:\: , \\ &&P^{\,(1)}_{\rm gg}(x) \: = \:
       4\, \* \casq  \*  \Big(
            2 \* \pgg(x) \* \Big[
            {67 \over 18}
          - \z2
          + \Hh(0,0)
          + 2 \* \Hh(1,0)
          + 2 \* \H(2)
          \Big]
          - 2 \* \pgg( - x) \* \Big[
            \z2
          + 2 \* \Hh(-1,0)
  \nonumber\\&&
          - \Hh(0,0)
          \Big]
          - {67 \over 9} \* \Big({1 \over x}-x^2\Big)
       - {44 \over 3} \*x^2\*\H(0)
          + (1+x) \* \Big[
         {11 \over 3} \* \H(0)
          + 8 \* \Hh(0,0)
       - {27 \over 2}
          \Big]
          + 27
          - 12 \* \H(0)
  \nonumber\\&&
          + \delta(1 - x) \* \Big[
            {8 \over 3}
          + 3 \* \z3
          \Big]
          \Big)
\:\: .\label{eq:Pgg1}
\eea
\normalsize
The corresponding three-loop contribution \cite{Vogt:2004mw} reads 
\small
\bea
  &&P^{\,(2)}_{\rm gg}(x) \: = \: 
       16\, \*  \cath  \*  \Big(
            \pgg(x) \* \Big[
            {245 \over 24} 
          - {67 \over 9} \* \z2
          + {11 \over 3} \* \z3
          - {3 \over 10} \* \z2^2
          - 4 \* \Hh(-3,0)
          + 6 \* \H(-2) \* \z2
          + 4 \* \Hhh(-2,-1,0)
  \nonumber\\&&
          + {11 \over 3} \* \Hh(-2,0)
          - 4 \* \Hhh(-2,0,0)
          - 4 \* \Hh(-2,2)
          + {1 \over 6} \* \H(0)
          - 7 \* \H(0) \* \z3
          + {67 \over 9} \* \Hh(0,0)
          - 8 \* \Hh(0,0) \* \z2
          + 4 \* \Hhhh(0,0,0,0)
          - 6 \* \H(1) \* \z3
  \nonumber\\&&
          - 4 \* \Hhh(1,-2,0)
          + {134 \over 9} \* \Hh(1,0)
          - 6 \* \Hh(1,0) \* \z2
          + {11 \over 6} \* \Hhh(1,0,0)
          + 8 \* \Hhhh(1,0,0,0)
          + 8 \* \Hhhh(1,1,0,0)
          + 8 \* \Hhh(1,2,0)
          + 8 \* \Hh(1,3)
  \nonumber\\&&
          + {134 \over 9} \* \H(2)
          - 4 \* \H(2) \* \z2
          + 10 \* \Hhh(2,0,0)
          + 8 \* \Hhh(2,1,0)
          + 8 \* \Hh(2,2)
          + {11 \over 6} \* \H(3)
          + 10 \* \Hh(3,0)
          + 8 \* \Hh(3,1)
          + 8 \* \H(4)
          \Big]
  \nonumber\\&&
          + \, \pgg(-x) \* \Big[
          - {67 \over 9} \* \z2
          + {11 \over 2} \* \z2^2
          - 4 \* \Hh(-3,0)
          + 16 \* \H(-2) \* \z2
          + 8 \* \Hhh(-2,-1,0)
          - 18 \* \Hhh(-2,0,0)
          - 12 \* \Hh(-2,2)
  \nonumber\\&&
          + 12 \* \H(-1) \* \z3
          + 8 \* \Hhh(-1,-2,0)
          - 16 \* \Hh(-1,-1) \* \z2
          + 24 \* \Hhhh(-1,-1,0,0)
          + 16 \* \Hhh(-1,-1,2)
          - {134 \over 9} \* \Hh(-1,0)
  \nonumber\\&&
          + 18 \* \Hh(-1,0) \* \z2
          - 16 \* \Hhhh(-1,0,0,0)
          - 4 \* \Hhh(-1,2,0)
          - 16 \* \Hh(-1,3)
          - {11 \over 6} \* \H(0) \* \z2
          - 5 \* \H(0) \* \z3
          + {67 \over 9} \* \Hh(0,0)
          - 8 \* \Hh(0,0) \* \z2
  \nonumber\\&&
          + 4 \* \Hhhh(0,0,0,0)
          + 2 \* \H(2) \* \z2
          + 2 \* \Hh(3,0)
          + 8 \* \H(4)
          \Big]
          + \Big({1 \over x}-x^2\Big) \* \Big[
            {16619 \over 162} 
          - {55 \over 2} \* \z3
          - {11 \over 2} \* \H(0) \* \z2
          - {413 \over 108} \* \H(1) 
  \nonumber\\&&
          - {11 \over 2} \* \H(1) \* \z2 
          - {67 \over 9} \* \Hh(1,0)
          + {33 \over 2} \* \Hhh(1,0,0) 
	     - {67 \over 9} \* \H(2)
          + {22 \over 3} \* \Hh(2,0)
          \Big]
          + 11 \* \Big({1 \over x}+x^2\Big) \* \Big[
          - {389 \over 198} \* \z2
	     - {2 \over 3} \* \Hh(-2,0)
  \nonumber\\&&
          - {1 \over 2} \* \H(-1) \* \z2
          + \Hhh(-1,-1,0) 
          - {523 \over 198} \* \Hh(-1,0) 
          + {8 \over 3} \* \Hhh(-1,0,0) 
          + \Hh(-1,2) 
          + {71 \over 54} \* \H(0)
          - {1 \over 6} \* \H(3)
          \Big]
          + x^2 \* \Big[
            {85 \over 6} \* \z2
  \nonumber\\&&
          + 33 \* \Hh(-2,0)
          + {6409 \over 108} \* \H(0)
          + 33 \* \H(0) \* \z2
          - {1249 \over 18} \* \Hh(0,0)
          - 44 \* \Hhh(0,0,0)
          - {44 \over 3} \* \Hh(2,0)
          - {110 \over 3} \* \H(3)
           \Big]
  \nonumber\\&&
          + (1-x) \* \Big[
          - {11317 \over 108} 
          -  4 \* \H(-3,0)
          - 4 \* \H(-2) \* \z2
          - 8 \* \H(-2,-1,0)
          - {19 \over 3} \* \H(-2,0)
          - 12 \* \H(-2,0,0)
	     - {263 \over 12} \* \H(0,0)
  \nonumber\\&&
          - {29 \over 3} \* \H(0,0,0)
          + {31 \over 36} \* \H(1)
          - {3 \over 2} \* \H(1) \* \z2
          + {27 \over 2} \* \H(1,0)
          - {25 \over 2} \* \H(1,0,0)
          \Big]
          + (1+x) \* \Big[
          - {329 \over 18} \* \z2
          + {11 \over 2} \* (1+x) \* \z3
  \nonumber\\&&
	     - {43 \over 5} \* \z2^2
          - {53 \over 2} \* \H(-1) \* \z2
          - 3 \* \Hhh(-1,-1,0)
          - {215 \over 6} \* \Hh(-1,0)
          + 38 \* \Hhh(-1,0,0)
          + 25 \* \Hh(-1,2)
	     + {4651 \over 216} \* \H(0)
          - 8 \* \H(0) \* \z3
  \nonumber\\&&
          + {27 \over 2} \* \H(0) \* \z2
          - 22 \* \Hh(0,0) \* \z2
	     - {158 \over 9} \* \H(2)
          - 4 \* \H(2) \* \z2
          + {29 \over 3} \* \Hh(2,0)
          + 10 \* \Hhh(2,0,0)
          - {43 \over 6} \* \H(3)
          + 16 \* \Hh(3,0)
          + 26 \* \H(4)
          \Big]
  \nonumber\\&&
          + {53 \over 6} \* \z2
          + 24 \* \z3
          + 2 \* \z2^2
          - 16 \* \Hh(-3,0)
          + 27 \* \Hh(-2,0)
          + {601 \over 12} \* \H(0)
          + {41 \over 3} \* \H(0) \* \z2
          - 16 \* \H(0) \* \z3
          - 29 \* \Hh(0,0)
  \nonumber\\&&
          - 4 \* \Hh(0,0) \* \z2
          - {40 \over 3} \* \Hhh(0,0,0)
          + 28 \* x \* \Hhhh(0,0,0,0)
          + 27 \* \H(2)
          - 24 \* \Hh(2,0)
          - 20 \* \H(3)
          + \delta(1 - x) \* \Big[
            {79 \over 32} 
          + {1 \over 6} \* \z2
  \nonumber\\&&
          + {67 \over 6} \* \z3
          + {11 \over 24} \* \z2^2
          - 5 \* \z5
          - \z2 \* \z3
          \Big]
          \Big)
\:\: ,\label{eq:Pgg2}
\eea
\normalsize
where, as in Eqs.~(\ref{eq:Pgg0}) and (\ref{eq:Pgg1}), all divergences for 
$x \ra 1$ are to be read as +-dis\-tributions.
Functions $H_{\vec{m}}(x)$ up to weight (number of indices) $2n$ occur 
at~N$^n$LO.

\vspace{1mm}
A {\sc Fortran} program for the harmonic polylogarithms up to weight four is 
available~\cite{Gehrmann:2001pz}. Nevertheless it is useful to have also more 
compact, if approximate representations of the three-loop splitting functions. 
Making use of the end-point behaviour discussed in the next section,
Eq.~(\ref{eq:Pgg2}) can be parametrized as~\cite{Vogt:2004mw}
\bea
\label{eq:Pgg-ap}
  P^{(2)}_{\rm gg}(x) & \cong & \mbox{} 
     + 2643.521\: {\cal D}_0 + 4425.894\: \delta(1-x) + 3589\: L_1
     - 20852\: + 3968\: x 
  \nn \\ & & \mbox{}
     - 3363\: x^2 + 4848\: x^3 + L_0 L_1\: ( 7305 + 8757\: L_0 )
     + 274.4\: L_0 
  \nn \\[0.5mm] & & \mbox{}
      - 7471\: L_0^2 + 72\: L_0^3 - 144\: L_0^4 + 14214\: x^{\,-1} 
      + 2675.8\: x^{\,-1} L_0
\eea

\vspace{-1.5mm}
\nin
where
\vspace{-4.5mm}

$$
  \DD_{\,0} \: \equiv \: 1/(1-x)_+ \: ,
  \quad L_1 \: \equiv \: \ln (1-x) \: ,
  \quad L_0 \: \equiv \: \ln x \:\: .
$$
This parametrization deviates from the exact expression (\ref{eq:Pgg2}) by 
less than 0.1\%, which should be perfectly sufficient for numerical 
applications. Note that the Mellin transform of Eq.~(\ref{eq:Pgg-ap}) can be 
readily continued to complex values of $N$ as required for the moment-space 
approach to the analysis of hard processes
\cite{Graudenz:1996sk}--\cite{Stratmann:2001pb}.

\section{Results and surprises for {\boldmath{$x \ra 1$}} and 
         {\boldmath{$x \ra 0$}}}
\setcounter{equation}{0}

\vspace{-0.5mm}
The end-point behaviour of the splitting functions is of particular interest. 
The leading contributions for $x \!\ra\! 1$ are related to the cusp anomalous 
dimensions and are thus relevant beyond the context of parton distributions. 
The perturbative stability at very small $x$, where potentially large 
$\,\ln^{k} x$ corrections occur, represents a much discussed 
topic directly relevant to analyses of collider processes.

\vspace{-0.5mm}
\subsection{The large-$x$ behaviour}

Up to N$^{n=2}$LO, at least, the diagonal \MSb-scheme splitting functions are 
given by
\beq
\label{eq:lx-diag}
  P^{\,(n)}_{{\rm aa}, x\ra 1}(x) \: = \:
     \frac{A_{n+1}^{\,\rm a}}{(1-x)_+} \: + \:  
     B_{n+1}^{\,\rm a}\: \delta(1-x) \: + \: 
     C_{n+1}^{\,\rm a}\: \ln (1-x) \: + \: {\cal O}(1) \:\: .
\eeq
In fact, the simple +-distribution $1/(1-x)_+$ constitutes the leading term to 
all orders --- in contrast to, for example, the coefficient functions in DIS 
which include terms $[(1-x)^{-1} \ln^{k} (1-x)]_+$ --- and its coefficients 
$A_{m}^{\,\rm a}$ form the perturbative expansion of the cusp anomalous 
dimensions of the respective Wilson lines~\cite{Korchemsky:1989si}. For the 
quark case the known coefficients read, in our normalization, 
\bea
\label{eq:cusp}
  A_1^{\,\rm q} & = & 4\, C_F \nn \\[0.1mm]
  A_2^{\,\rm q} & = & 8\, C_F \left[ \left( \frac{67}{18} - \zeta_2^{} \right)
                C_A - \frac{5}{9}\,\nf \right] \nn \\[0.1mm]
  A_3^{\,\rm q} & = & 
     16\, C_F C_A^{\,2} \, \left[ \frac{245}{24} - \frac{67}{9}\: \z2
         + \frac{11}{6}\:\z3 + \frac{11}{5}\:\z2^{\!\! 2} \right]
   +  16\, C_F^{\,2} \nf\, \left[ -  \frac{55}{24}  + 2\:\z3 \right] 
   \quad \nn \\ & & \mbox{}
   +  16\, C_F C_A \nf\, \left[ - \frac{209}{108}
         + \frac{10}{9}\:\z2 - \frac{7}{3}\:\z3 \right]
   \, +\,  16\, C_F \n2f \left[ - \frac{1}{27}\,\right] \:\: .
\eea
The coefficients $A_{m}^{\,\rm g}$ are obtained from Eq.~(\ref{eq:cusp}) by
multiplication with $C_A/C_F$. Note that the $\nf$-independent parts of 
$A_{m}^{\,\rm a}$ consist of only one (the maximally non-abelian) colour 
factor as also predicted in ref.~\cite{Korchemsky:1989si}.
The $\nf\! =\! 0$ part of $A_3^{\,\rm q}$ and the complete $A_3^{\,\rm g}$ are 
new results of refs.~\cite{Moch:2004pa} and \cite{Vogt:2004mw}, respectively. 
Two computations of the $\nf$-contribution to $A_3^{\,\rm q}$ were performed 
two years ago~\cite{Moch:2002sn,Berger:2002sv}, and its $\n2f$-part was already 
obtained in ref.~\cite{Gracey:1994nn}. 
$A_3^{\,\rm q}$ in Eq.~(\ref{eq:cusp}) agrees with the previous numerical 
estimate~\cite{Vogt:2000ci} which has been widely used in soft-gluon
resummation analyses.

\vspace{1mm}
It is also interesting that at three loops, as in the previous order, only a 
single logarithm occurs in Eq.~(\ref{eq:lx-diag}). Furthermore it turns out 
that there is an unexpected relation between the corresponding coefficients 
$C_{n+1}^{\,\rm a}$ and the $A_{n}^{\,\rm a}$, viz
\cite{Moch:2004pa,Vogt:2004mw}
\beq
\label{eq:lx-log}
  C_1^{\,\rm a} \: = \: 0 \: , \quad 
  C_2^{\,\rm a} \: = \: (A_1^{\,\rm a})^2 \: , \quad
  C_3^{\,\rm a} \: = \: 2\:\! A_1^{\,\rm a}  A_2^{\,\rm a} \:\: .
\eeq
The logarithmic structure of Eq.~(\ref{eq:lx-diag}) and especially the relation
(\ref{eq:lx-log}) call for an explanation and, possibly, a higher-order 
generalization.

\vspace{1mm}
The large-$x$ limit of the quark-gluon and gluon-quark splitting functions 
reads 
\beq
   \textstyle{
   P^{\,(n)}_{ab,x\ra 1}(x) \: = \: \sum_{i=0}^{2n-1} D_{n,i}^{\,\rm ab}\,
   \ln^{2n-i} (1\! -\! x) \: + \: {\cal O}(1) \:\: .}
\eeq
See ref.~\cite{Vogt:2004mw} for the three-loop coefficients $D_{2,i}^{\,\rm ab}
$.  Except for $D_{2,3}^{\,\rm gq}$ these coefficients vanish for the choice
$C_A = C_F = \nf$ leading to a $\,{\cal N}=1$ supersymmetric theory.

\subsection{The small-$x$ behaviour}

We start with the flavour non-singlet contributions which, in the present 
unpolarized case, are practically far less important then the singlet parts. 
However, the non-singlet small-$x$ expansion includes two additional powers of 
$\ln x$ per order, i.e., terms up to $\ln^{\, 2n} x$ occur at N$^n$LO. Thus the 
three-loop splitting functions
\beq
\label{eq:sx-ns}
  P^{(2)i=\pm,\rm s}_{x\ra 0}(x) \: = \: D_0^{\, i} \ln^4 x \:
   + \: \ldots \:
  + \: D_3^{\, i} \ln x \: + \: {\cal O}(1) 
\eeq
form a presently unique theoretical laboratory for studying the relative size 
of as many as four small-$x$ logarithms as obtained from a complete (all-$x$)
calculation.
 
\vspace{1mm}
The numerical QCD values of the coefficients for $\,i=-\,$ in 
Eq.~(\ref{eq:sx-ns}) read 
\bea
  D_0^- &\:\cong\: & 1.4321   \nn \\
  D_1^- &\:\cong\: & 35.556 - 3.1605\: \nf   \nn \\
  D_2^- &\:\cong\: & 399.21 - 39.704\: \nf + 0.5926\: n_{\! f}^{\, 2} \nn \\
  D_3^- &\:\cong\: & 1465.9 - 172.69\: \nf + 4.3457\: n_{\! f}^{\, 2} \:\: .
\eea
See ref.~\cite{Moch:2004pa} for the analytic expressions and the similar case 
$i=+$. In both cases the leading coefficients $D_0$ have been correctly 
predicted~\cite{Blumlein:1996jp} on the basis of the resummation in
ref.~\cite{Kirschner:1983di}.  Note that the expansion (\ref{eq:sx-ns})
alternates for $i=\pm$, and that the coefficients $D_k$ rise sharply with~$k$.
For $\nf = 4$ and $\,i=-\,$ shown in Fig.~2, the modulus of the $\ln^4 x$ 
($\ln^3 x$) contribution is twice as large as that of the next term, $\ln^3 x$ 
($\ln^2 x$), only at extremely small $x$-values, $x \lsim 10^{-14}$ 
($3 \cdot 10^{-10}$). 
 
\vspace{1mm}
The numerical situation is rather different for the $d_{abc}d_{abc}$ 
contribution $i=\rm s$~\cite{Moch:2004pa},
\bea
  D_0^{\, \rm s} & \,\cong\, & + 1.4815\: \nf  \:\: , \quad
  D_1^{\, \rm s}   \,\cong\,   - 2.9630\: \nf  \nn \\
  D_2^{\, \rm s} & \,\cong\, & + 6.8918\: \nf  \:\: , \quad
  D_3^{\, \rm s}   \,\cong\,   + 178.03\: \nf  \:\: .
\eea
Here the leading small-$x$ terms do indeed provide a reasonable approximation,
see Fig.~2. Note that the existence of a leading ($\ln^4 x$) $d_{abc}d_{abc}$ 
contribution for $i=\rm s$ is rather surprising. In fact, the presence of a
leading small-$x$ logarithm in a term unpredictable from lower-order structures 
appears to call into question the very concept of the small-$x$ resummation of 
the double logarithms $\alpha_{\rm s}^{n+1} \ln^{\, 2n} x$.

\vspace{1mm}
The leading small-$x$ terms of the three-loop singlet splitting functions
are 
\beq
\label{eq:sx-sg}
  P^{(2)}_{{\rm ab},x\ra 0}(x) \: = \: E_1^{\,\rm ab}\: \frac{\ln x}{x}
  \: +\: E_2^{\,\rm ab}\: \frac{1}{x} \: +\: {\cal O}( \ln^{\,4} x) \:\: .
\eeq
In general, $x^{-1} \ln^{\, k} x$ contributions with $k\leq n$ occur in 
$P_{\rm gq}$ and $P_{\rm gg}$ at N$^n$LO. The highest of these terms ($k=n$) have, 
however, vanishing coefficients for $n\!=\!1,\,2$ as predicted by the 
leading-logarithmic BFKL equation~\cite{Kuraev:1977fs,Balitsky:1978ic}, see 
also ref.~\cite{Jaroszewicz:1982gr}.

\vspace{1mm}
In QCD the numerical values of the coefficients in Eq.~(\ref{eq:sx-sg}) are 
given by 
\bea
  E_1^{\,\rm qq} \:\cong\: - 132.74  \, \nf   \:\:\quad\qquad &,& \:\: 
  E_2^{\,\rm qq} \:\cong\: - 506.00 \, \nf + 3.1605\, \n2f \nn \\
  E_1^{\,\rm qg} \:\cong\: - 298.67 \, \nf   \:\:\quad\qquad &,& \:\:
  E_2^{\,\rm qg} \:\cong\: - 1268.3 \, \nf + 4.5761\, \n2f \nn \\
%
  E_1^{\,\rm gq} \:\cong\: 1189.3 + 71.083\, \nf \: &,& \:\:
  E_2^{\,\rm gq} \:\cong\: 6163.1 - 46.408\, \nf
    - 2.3704\, \n2f \quad \nn \\
  E_1^{\,\rm gg} \:\cong\: 2675.9 + 157.27\, \nf \: &,& \:\:
  E_2^{\,\rm gg} \:\cong\: 14214. + 182.96\, \nf - 2.7984\, \n2f \:\: .
\eea
The analytical results can be found in ref.~\cite{Vogt:2004mw}. The 
coefficients $E_1^{\,\rm qa}$ and $E_1^{\,\rm gg}$ agree with those derived
from the small-$x$ resummation in refs.~\cite{Catani:1994sq} and 
\cite{Fadin:1998py}, respectively, after transforming the latter result to the 
\MSb\ scheme \cite{vanNeerven:2000uj}. $E_1^{\,\rm gq}$ was unknown before  
ref.~\cite{Vogt:2004mw}.
For $\nf = 3 \ldots 5\,$ the ratios $\,E_2^{\,\rm ab}/E_1^{\,\rm ab\,}$ are
$ 3.7 \ldots 4.7$. Thus the corrections due to the non-logarithmic $1/x$
terms amount to less than 50\% only at $x \lsim 10^{-4}$. 


\begin{figure}[htbp]
\vspace{-4mm}
\label{fig2}
\centerline{\epsfig{file=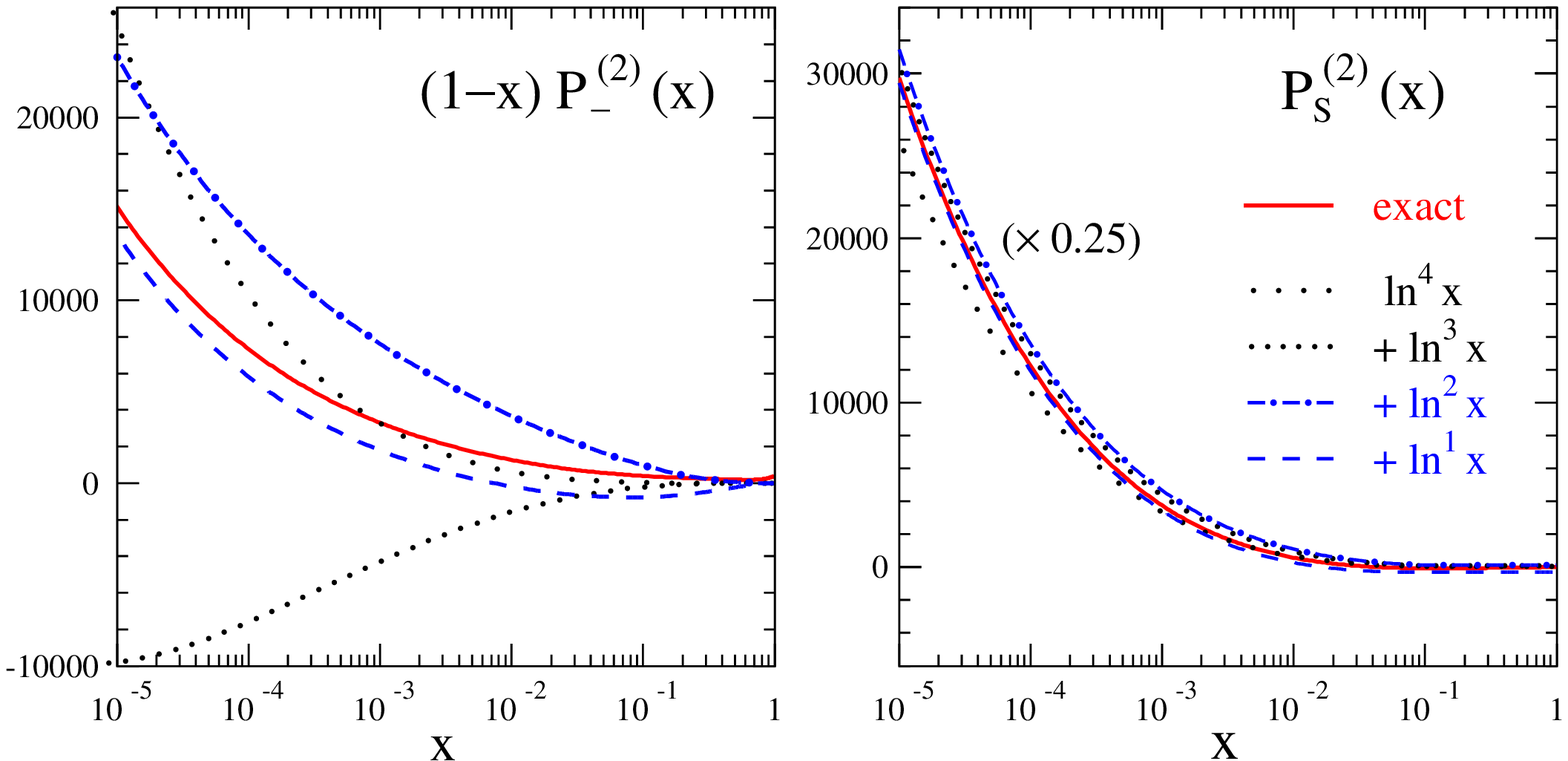,width=10.8cm,angle=0}}
\centerline{\epsfig{file=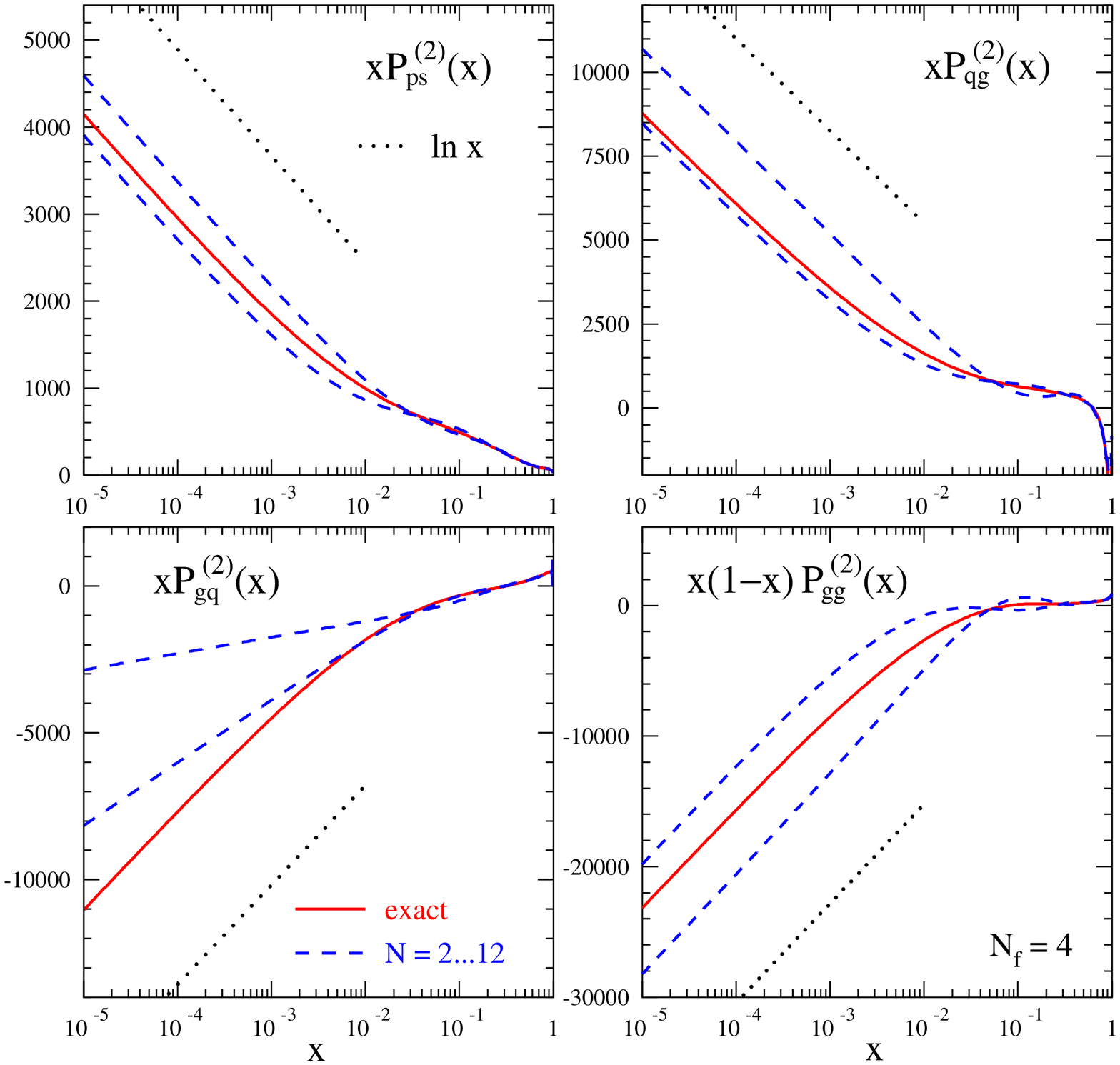,width=10.8cm,angle=0}}
\vspace{-2mm}
\caption{The small-$x$ behaviour of non-singlet (upper row) and singlet
 NNLO splitting functions for $\nf\! =\! 4$. Also shown are the (successive) 
 approximations (\protect\ref{eq:sx-ns}) and (\protect\ref{eq:sx-sg}) by the 
 leading small-$x$ logarithms and, for the singlet cases, the errors bands
 \protect\cite{vanNeerven:2000wp} used in the analyses of refs.~\protect\cite
 {Martin:2002dr,Alekhin:2002fv}.}
\vspace*{-2mm}
\end{figure}

\section{The size of the corrections}
\setcounter{equation}{0}

Finally we discuss the numerical effects of our new contributions $P^{\,(2)}$ 
to Eq.~(\ref{eq:P-exp}). For brevity, we confine ourselves to four quark 
flavours and 
a typical scale $\mu^2 \simeq 30 \ldots 50 \mbox{ GeV}^2$. We employ an 
order-independent value of the strong coupling, 
\beq
\label{eq:as-ref}
  \as(\mu_0^2, \nf=4 ) = 0.2 \:\: , 
\eeq
facilitating a direct comparison of the \MSb\ evolution kernels at LO, NLO and 
NNLO.

\subsection{N-space$\,$: anomalous dimensions}

The singlet anomalous dimensions $\gamma^{}_{\:\!\rm f\:\!f'}$ are displayed in 
Fig.~3 for the standard choice $\mu_r = \mu_f$ of the renormalization scale 
tacitly made already in sections 3 and 4.
The NNLO corrections are markedly smaller than the NLO contributions. 
For the choice (\ref{eq:as-ref}) they amount, at $N\! >\! 2$, to less than 2\%
and 1\% for the large diagonal quantities $\gamma_{\:\!\rm qq}$ and 
$\gamma_{\:\!\rm gg}$, respectively, while for the much smaller off-diagonal 
anomalous dimensions $\gamma_{\:\!\rm qg}$ and $\gamma_{\:\!\rm gq}$ values of 
up to 6\% and 4\% are reached. Also shown in Fig.~3 is the pure-singlet 
contribution $\gamma_{\:\!\rm ps}$ defined by Eq.~(\ref{eq:Pqq}). At 
$N > 2$ this quantity receives very large relative (but tiny absolute) NNLO 
corrections.

\begin{figure}[htb]
\label{fig3}
\vspace{-2mm}
\centerline{\epsfig{file=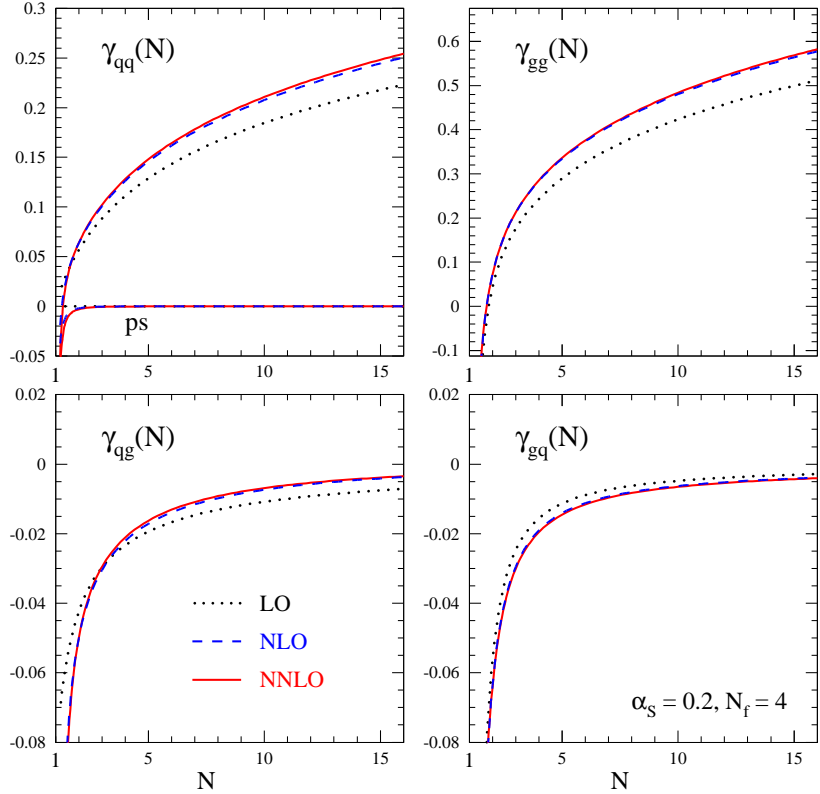,width=11cm,angle=0}}
\vspace{-2mm}
\caption{The perturbative expansion of the singlet anomalous dimensions 
 $\gamma^{}_{\rm f\:\!f'}$ up to NNLO.}
\vspace*{-2mm}
\end{figure}

\subsection{Scale derivatives of $x$-space parton distributions}

In Figs.~4 and 5 we show the logarithmic derivatives $\dot{f_k} = d\ln f_k /
d\ln \mu_{\! f}^2$ for the sufficiently realistic -- and like Eq.\ 
(\ref{eq:as-ref}) order-independent -- model distributions
\bea
\label{eq:f-ref}
  xq_{\rm ns}^{}(x,\mu_{0}^{\,2}) & \: = \: & 
  x^{\, 0.5}\, (1-x)^3 \nn \\
  xq_{\rm s}^{}(x,\mu_{0}^{\,2}) \: &\: = \: &
  0.6\: x^{\, -0.3}\, (1-x)^{3.5}\, (1 + 5.0\: x^{\, 0.8\,}) \nn \\
  xg (x,\mu_{0}^{\,2}) \:\: &\: = \: &
  1.6\: x^{\, -0.3}\, (1-x)^{4.5}\, (1 - 0.6\: x^{\, 0.3\,}) \:\: .
\eea
At large $x$ the NNLO corrections to the non-singlet evolution illustrated in
Fig.~4 are very similar for all three cases (\ref{eq:q-ns}).
They amount to 2\% or less for $x\geq 0.2$, thus being smaller than the NLO
contributions by a factor of about eight. For $q_{\,\rm ns}^{\,-}$ the same
suppression is also found in the region $10^{-5}\lsim x \lsim 10^{-2}$.
The NNLO effects are even smaller for $q_{\,\rm ns}^{\,+}$ (not shown) at small
$x$, but considerably larger for $q_{\,\rm ns}^{\,\rm v}$ at $x< 10^{-3}$ due
to the additional effect of the new quantity $P_{\rm ns}^{\,\rm s}$ in 
Eq.~(\ref{eq:p-ns}).

\begin{figure}[htbp]
\label{fig4}
\vspace{-2mm}
\centerline{\epsfig{file=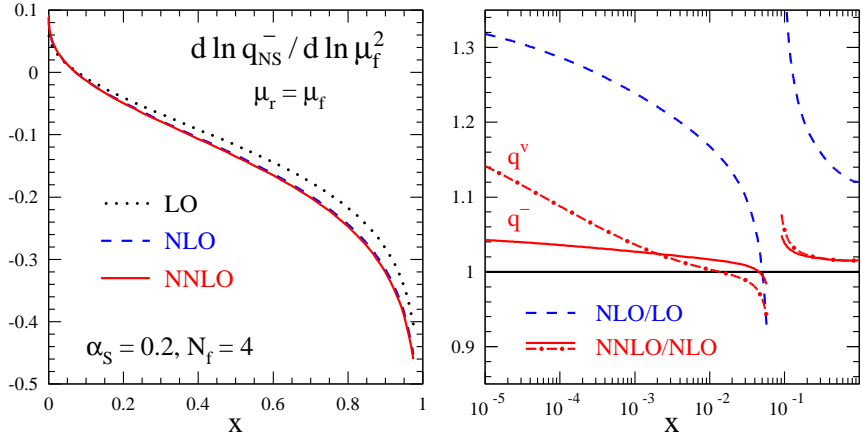,width=11.5cm,angle=0}}
\vspace{-3mm}
\caption{The perturbative expansion up to NNLO of the factorization-scale 
 derivatives $\dot{q}_{\,\rm ns}^{\, -,\rm v}$ for the initial conditions 
 (\ref{eq:as-ref}), (\ref{eq:f-ref}) and the standard choice $\mu_r = \mu_f$ of 
 the renormalization scale.}
\vspace*{-3mm}
\end{figure}

Also for the singlet quark distribution (upper row of Fig.~5) the ratio of the
NLO and NNLO corrections is about eight over the full $x$-range. However, at 
small$~x$ -- where $P_{\rm qg} \ast g$ dominates in Eq.~(\ref{eq:ev-sg}) -- 
the LO results are anomalously small as $P_{\rm qq}$ and $P_{\rm qg}$, unlike 
at higher orders, do not include $1/x$ terms at first order.
The situation is quite different for the evolution of the gluon density
(dominated by $P_{\rm gg} \ast g$ at all $x$). Here the NLO contributions 
appear atypically small at low $x$, cf.~the remark below Eq.~(\ref{eq:sx-sg}).
Thus the ratio of the NNLO and NLO corrections is rather large here,
despite the former amounting to only 3\% for $x$ as low as $10^{\,-4}$.

\begin{figure}[htbp]
\label{fig5}
\vspace{1mm}
\centerline{\epsfig{file=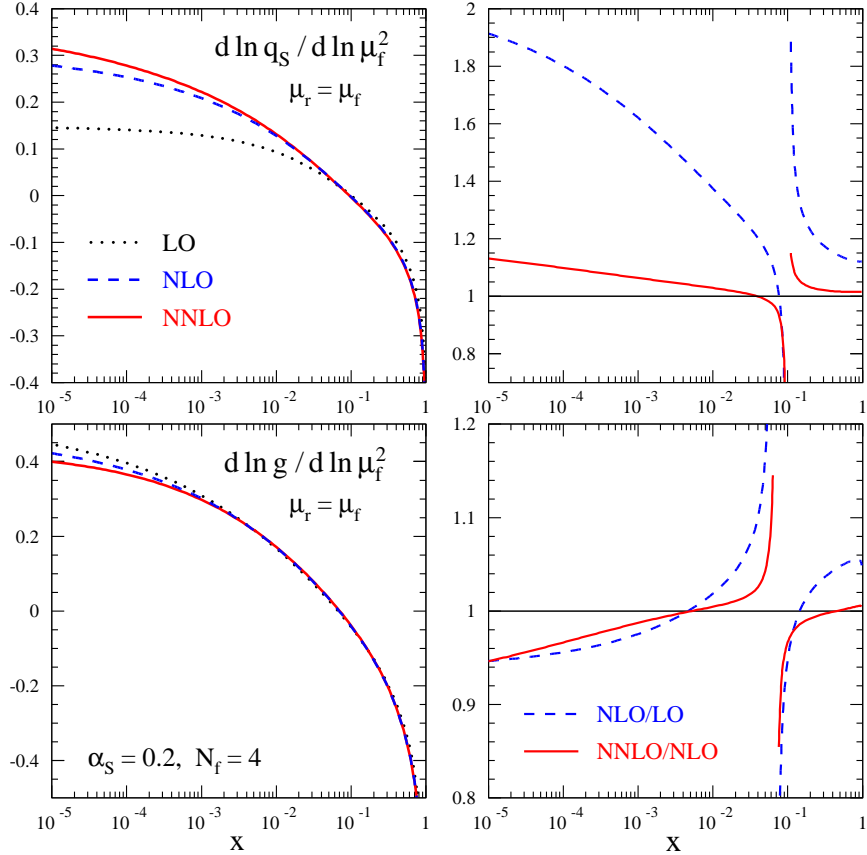,width=11.5cm,angle=0}}
\vspace{-2mm}
\caption{As Fig.~4, but for the singlet quark and gluon distributions 
 $q_{\:\!\rm s}$ and $g$. The spikes close $x = 0.1$ in the right plots reflect 
 the zeros of $\dot{f}_k$ and do not constitute large absolute corrections.}
\vspace*{-3mm}
\end{figure}

\vspace{1mm}
It is also interesting to consider the stability of the above results under
variations of the renormalization scale $\mu_r$. For $\mu_r\neq \mu_f$ the 
perturbative expansion of the splitting functions up to NNLO reads, with
$L_{fr} \,\equiv\, \ln (\mu_f^{\,2}/\mu_r^2)$,
\bea
  P & \: = \: & \quad
    a_{\rm s}(\mu_{r}^{\,2}) \: P^{(0)}  \:\: + \:\:
    a_{\rm s}^{\,2}(\mu_{r}^{\,2}) \, \big( P^{(1)} - \beta_0\, P^{(0)} 
      L_{fr} \big) \: \nn \\
 & & \mbox{}\!\!\!  + \:  
    a_{\rm s}^{\,3}(\mu_{r}^{\,2}) \, \big( P^{(2)} - \{ \beta_1 P^{(0)} 
      + 2\beta_0\, P^{(1)} \} L_{fr} + \beta_0^2\, P^{(0)} L_{fr}^2 
      \big) \:\: .
\eea
Here $a_{\rm s}(\mu_{\,r}^{\,2})$ is obtained at N$^n$LO from the value 
(\ref{eq:as-ref}) at the scale $\mu_0^2$ by solving 
\beq
\label{eq:as-run}
 \frac{da_{\rm s}}{d \ln \mu_r^2} \: = \: \beta (a_{\rm s}) \: = \: 
 - \sum_{l=0}^{n} \, a_{\rm s}^{\,k+2} \beta_k \:\: .
\eeq
The \MSb\ expansion coefficients $\beta_k$ are presently known up to $k=3$
\cite{Caswell:1974gg}--\cite{vanRitbergen:1997va}.

\vspace{1mm}
The dependence of the above results on $\mu_r$ can be found in Fig.~8 
of ref.~\cite{Moch:2004pa} and Figs.~9 and 10 of ref.~\cite{Vogt:2004mw} for 
selected values of $x$. 
Here we only show, in Fig.~6, the relative scale uncertainties $\Delta 
\dot{f_k}$ of the average $\mu_f$-derivatives as conventionally estimated by 
varying $\mu_{r}$ up to a factor of two with respect to~$\mu_f$,
\beq
\label{eq:mr-rel}
 \Delta \dot{f} \: = \:
 \frac{\max\, \dot{f} - \min\, \dot{f}}
 { 2\, |\, {\rm average}\, \dot{f} | } \:\: , \quad
 \mu_r^2 = \frac{1}{4}\, \mu_{\! f}^2 \ldots\, 4\, \mu_{\! f}^2 \:\: .
\eeq
For the valence, (singlet-quark, gluon) distributions, these uncertainty 
estimates amount to 3\% (3\%, 1\%) or less at $x > 5\cdot 10^{-4\,}$ 
($5\cdot 10^{-3\,}$, $3\cdot 10^{-4\,}$), an improvement by more than a factor 
of three with respect to the corresponding NLO results. 

\begin{figure}[htb]
\vspace{1mm}
\label{fig6}
\centerline{\epsfig{file=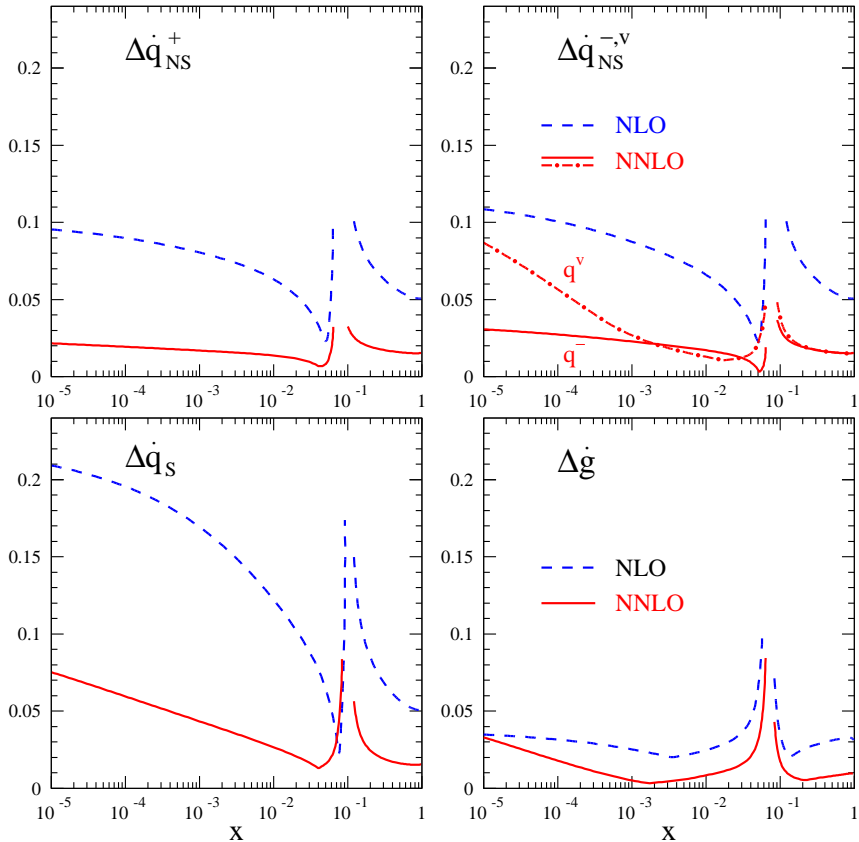,width=11.5cm,angle=0}}
\vspace{-2mm}
\caption{The renormalization scale uncertainties of the NLO and NNLO
 predictions for the evolution of the non-singlet and singlet 
 distributions as estimated by the quantities $\Delta \dot{f}_k$ 
 in Eq.~(\ref{eq:mr-rel}).}
\vspace{-3mm}
\end{figure}

\section{Summary and outlook}

\vspace{-1mm}
We have calculated the complete third-order splitting functions for the 
evolution of unpolarized parton distributions in perturbative QCD.
The computation is performed in Mellin-$N$ space and follows the previous 
fixed-$N$ computations~\cite{Larin:1994vu}--\cite{Retey:2000nq} inasmuch as we 
compute the partonic structure functions in deep-inelastic scattering at 
even or odd $N$. Our calculation, however, provides the complete $N$-dependence 
from which the splitting functions in Bjorken-$x$ space can be uniquely 
reconstructed.

\vspace{1mm}
A salient feature of our approach is that it facilitates very efficient checks
of our elaborate schemes for the reduction of the required three-loop integrals
by using the {\sc Mincer} program \cite{Gorishnii:1989gt,Larin:1991fz}.
By keeping terms of order $\varepsilon^0$ in dimensional regularization 
throughout the calculation, we have also obtained the third-order coefficient 
functions for the structure functions $F_2$ and $F_L$ in electromagnetic and 
for $F_3$ in charged-current DIS \cite{MVV5}. 
The present method can be used to generalize our fixed-$N$ three-loop 
calculation of the photon structure \cite{Moch:2001im} to all $N$. It should
also be possible to obtain the polarized NNLO splitting function in this
manner.
 
Our results completely agree with all partial results available in the
literature for fixed moments~\cite{Larin:1994vu}--\cite{Retey:2000nq}, 
large-$\nf$ limits~\cite{Gracey:1994nn,Bennett:1997ch}, small-$x$ behaviour
\cite{Catani:1994sq}--\cite{Fadin:1998py} and large-$x$ structure \cite
{Berger:2002sv,Korchemsky:1989si}. With the (relatively unimportant) exception 
of $P_{\rm gq}$ shown above and $P_{\rm ns}^{\,\rm s}$ they also fully agree 
with the uncertainty bands of ref.~\cite{vanNeerven:2000wp} used in provisional 
NNLO analyses~\cite{Martin:2002dr,Alekhin:2002fv}. Those analyses thus remain 
valid. The results do, however, exhibit some unexpected features, most notably
a suggestive relation between large-$x$ coefficients and the presence of a 
leading small-$x$ term in the new $d_{abc}d_{abc}$ contribution $P_{\rm ns}
^{\,\rm s}$ to the splitting function for the valence distribution. 

\vspace{1mm}
The effect of the three-loop (NNLO) corrections on the evolution of the parton
densities is small at $x \gsim 10^{-3}$. For $\as = 0.2$, for example, 
both the corrections and the $\mu_r$ variation amount to less than 2\% at large
$x$; and the NNLO effects are about eight times smaller than the NLO 
contributions, implying that the evolution is perturbatively stable down to 
rather low scales. 
For $x\! <\! 10^{-3}$ the corrections increase with decreasing $x$. As the 
knowledge of the leading small-$x$ terms is not sufficient, further 
improvements in this region would require considerable efforts, including at 
least an extension of the {\sc Mincer} program 
\cite{Gorishnii:1989gt,Larin:1991fz} to four loops.



\vspace{-2mm}
\begin{thebibliography}{99}
\vspace{-1mm}

\bibitem{Moch:2004pa}
S. Moch, J.A.M. Vermaseren and A. Vogt,
  Nucl.~Phys.~B688 (2004) 101, \mbox{hep-ph/0403192} 

\bibitem{Vogt:2004mw}
A. Vogt, S. Moch and J.A.M. Vermaseren, 
  Nucl.~Phys.~B691 (2004) 129, \mbox{hep-ph/0404111} 

\bibitem{Gross:1973rr}
D.J. Gross and F. Wilczek,
  Phys.~Rev.~D8 (1973) 3633

\bibitem{Georgi:1974sr}
H. Georgi and H.D. Politzer,
  Phys.~Rev.~D9 (1974) 416

\bibitem{Altarelli:1977zs}
G. Altarelli and G. Parisi,
  Nucl.~Phys.~B126 (1977) 298

\bibitem{Floratos:1977au}
E.G. Floratos, D.A. Ross and C.T. Sachrajda,
  Nucl.~Phys.~B129 (1977) 66

\bibitem{Floratos:1979ny}
E.G. Floratos, D.A. Ross and C.T. Sachrajda,
  Nucl.~Phys.~B152 (1979) 493

\bibitem{Curci:1980uw}
G. Curci, W. Furmanski and R. Petronzio,
  Nucl.~Phys.~B175 (1980) 27

\bibitem{Furmanski:1980cm}
W. Furmanski and R. Petronzio,
  Phys.~Lett.~97B (1980) 437

\bibitem{Floratos:1981hs}
E.G. Floratos, C. Kounnas and R. Lacaze,
  Nucl.~Phys.~B192 (1981) 417

\bibitem{Hamberg:1992qt}
R. Hamberg and W.L. van Neerven,
  Nucl.~Phys.~B379 (1992) 143

\bibitem{Larin:1994vu}
S. Larin, T.~van Ritbergen, and J. Vermaseren,
  Nucl.~Phys.~B427 (1994) 40

\bibitem{Larin:1997wd}
S.~Larin, P.~Nogueira, T.~van Ritbergen and J.~Vermaseren,
  Nucl.~Phys.~B492 (1997) 338, hep-ph/9605317

\bibitem{Retey:2000nq}
A. Retey and J. Vermaseren,
  Nucl.~Phys.~B604 (2001) 281, hep-ph/0007294

\bibitem{Gracey:1994nn}
J.A. Gracey,
  Phys.~Lett.~B322 (1994) 141, hep-ph/9401214

\bibitem{Bennett:1997ch}
J.F. Bennett and J.A. Gracey,
  Nucl.~Phys.~B517 (1998) 241, hep-ph/9710364

\bibitem{Catani:1994sq}
S. Catani and F. Hautmann,
  Nucl.~Phys.~B427 (1994) 475, hep-ph/9405388

\bibitem{Blumlein:1996jp}
J. Bl\"umlein and A. Vogt,
  Phys.~Lett.~B370 (1996) 149, hep-ph/9510410

\bibitem{Fadin:1998py}
V.S. Fadin and L.N. Lipatov,
  Phys.~Lett.~B429 (1998) 127, hep-ph/9802290

\bibitem{Moch:2002sn}
S. Moch, J.A.M. Vermaseren and A. Vogt, 
  Nucl.~Phys.~B646 (2002) 181, \mbox{hep-ph/0209100}

\bibitem{Vermaseren:2002rn}
J. Vermaseren, S. Moch and A. Vogt,
  Nucl.~Phys.~Proc.~Suppl.~116 (2003) 100, hep-ph/0211296

\bibitem{Berger:2002sv}
C.F. Berger,
  Phys.~Rev.~D66 (2002) 116002, hep-ph/0209107

\bibitem{MVV5}
J.A.M. Vermaseren, A. Vogt and S. Moch,
  in preparation

\bibitem{Gorishnii:1989gt}
S.G. Gorishnii et~al.,
  Comput.~Phys.~Commun.~55 (1989) 381

\bibitem{Larin:1991fz}
S.A. Larin, F.V. Tkachev and J.A.M. Vermaseren,
  NIKHEF-H-91-18

\bibitem{Larin:1991tj}
S. Larin and J. Vermaseren,
  Phys.~Lett.~B259 (1991) 345

\bibitem{vanNeerven:1991nn}
W.L. van Neerven and E.B. Zijlstra,
  Phys.~Lett.~B272 (1991) 127

\bibitem{Zijlstra:1991qc}
E.B. Zijlstra and W.L. van Neerven,
  Phys.~Lett.~B273 (1991) 476

\bibitem{Zijlstra:1992kj}
E.B. Zijlstra and W.L. van Neerven,
  Phys.~Lett.~B297 (1992) 377

\bibitem{Zijlstra:1992qd}
E.B. Zijlstra and W.L. van Neerven,
  Nucl.~Phys.~B383 (1992) 525

\bibitem{Ybook}
F.J. Yndur\'ain, {\it The Theory of Quark and Gluon Interactions}, 3rd edition,
 \\ (Springer 1999) and references therein.

\bibitem{ZG2a}
H.~Kluberg-Stern and J.B. Zuber,
  Phys.\ Rev.~D12 (1975) 467

\bibitem{ZG2b}
J.C. Collins, A.~Duncan, and S.D. Joglekar,
  Phys.~Rev.~D16 (1977) 438

\bibitem{Nogueira:1991ex}
P. Nogueira,
  J. Comput.~Phys.~105 (1993) 279

\bibitem{Vermaseren:2000nd}
J.A.M. Vermaseren,
  math-ph/0010025

\bibitem{Vermaseren:2002rp}
J.A.M. Vermaseren,
  Nucl.~Phys.~Proc.~Suppl.~116 (2003) 343, hep-ph/0211297

\bibitem{Vermaseren:1998uu}
J.A.M. Vermaseren,
  Int.~J. Mod.~Phys.~A14 (1999) 2037, hep-ph/9806280

\bibitem{Remiddi:1999ew}
E. Remiddi and J.A.M. Vermaseren,
  Int.~J.~Mod.~Phys.~A15 (2000) 725, hep-ph/9905237

\bibitem{Moch:1999eb}
S. Moch and J. Vermaseren,
  Nucl.~Phys.~B573 (2000) 853, hep-ph/9912355

\bibitem{Gehrmann:2001pz}
T. Gehrmann and E. Remiddi,
  Comput.~Phys.~Commun.~141 (2001) 296, hep-ph/0107173

\bibitem{Graudenz:1996sk}
Ch.\ Berger, D. Graudenz, M. Hampel and A. Vogt,
  Z. Phys.~C70 (1996) 77, hep-ph/9506333

\bibitem{Kosower:1998vj}
D. A. Kosower,
  Nucl.~Phys.~B520 (1998) 263, hep-ph/9708392

\bibitem{Stratmann:2001pb}
M. Stratmann and W. Vogelsang,
  Phys.~Rev.~D64 (2001) 114007, \\ hep-ph/0107064

\bibitem{Korchemsky:1989si}
G.~P.~Korchemsky,
  Mod.~Phys.~Lett.~A4 (1989) 1257

\bibitem{Vogt:2000ci}
A. Vogt,
  Phys.~Lett.~B497 (2001) 228, hep-ph/0010146

\bibitem{Kirschner:1983di}
R. Kirschner and L.N. Lipatov,
  Nucl.~Phys.~B213 (1983) 122

\bibitem{Kuraev:1977fs}
E.A.~Kuraev, L.N.~Lipatov and V.S.~Fadin,
Sov.\ Phys.\ JETP 45 (1977) 199

\bibitem{Balitsky:1978ic}
I.I.~Balitsky and L.N.~Lipatov,
Sov.\ J.\ Nucl.\ Phys.\ 28 (1978) 822

\bibitem{Jaroszewicz:1982gr}
T.~Jaroszewicz,
Phys.\ Lett.\ B116 (1982) 291.

\bibitem{vanNeerven:2000uj}
W.L. van Neerven and A. Vogt,
  Nucl.~Phys.~B588 (2000) 345, hep-ph/0006154

\bibitem{vanNeerven:2000wp}
W.L. van Neerven and A. Vogt,
  Phys.~Lett.~B490 (2000) 111, hep-ph/0007362

\bibitem{Martin:2002dr}
A.D.~Martin, R.G.~Roberts, W.J.~Stirling and R.S.~Thorne,
  Phys.~Lett.~B531 (2002) 216, hep-ph/0201127

\bibitem{Alekhin:2002fv}
S.~Alekhin,
  Phys.~Rev.~D68 (2003) 014002, hep-ph/0211096

\bibitem{Caswell:1974gg}
W.E. Caswell,
  Phys.~Rev.~Lett.~33 (1974) 244

\bibitem{Jones:1974mm}
D.R.T. Jones,
  Nucl.~Phys.~B75 (1974) 531

\bibitem{Tarasov:1980au}
O.V. Tarasov, A.A. Vladimirov, and A.Y. Zharkov,
  Phys.~Lett.~93B (1980) 429

\bibitem{Larin:1993tp}
S. Larin and J. Vermaseren,
  Phys.~Lett.~B303 (1993) 334, hep-ph/9302208

\bibitem{vanRitbergen:1997va}
T.~van Ritbergen, J.~Vermaseren and S.~Larin,
  Phys.\ Lett.\ B400 (1997) 379, hep-ph/9701390

\bibitem{Moch:2001im}
S.~Moch, J.A.M.~Vermaseren and A.~Vogt,
  Nucl.\ Phys.\ B621 (2002) 413, \mbox{hep-ph/0110331}


\end{thebibliography}
\end{document}